\documentclass[useAMS,usenatbib]{mn2e}
\usepackage{epsfig}
\usepackage{epstopdf}
\usepackage{graphicx}
\usepackage{amssymb}
\usepackage{amsmath}
\usepackage{graphicx}
\usepackage{subcaption}
\usepackage[utf8]{inputenc}
\usepackage[export]{adjustbox}
\usepackage{wrapfig}
\usepackage{textcomp}
\usepackage[usenames,dvipsnames]{color}
\usepackage{gensymb}
\usepackage{booktabs}
\usepackage{amsmath}
\usepackage{lscape}
\usepackage[english]{babel}
\usepackage{graphicx}
\usepackage{float}
\usepackage{fixltx2e}

\newcommand{\Msol}{\ensuremath{\mathrm{M}_{\odot}}}

\title[Galaxy Clustering in VIDEO]{The galaxy-halo connection in the VIDEO Survey at $0.5<z<1.7$}
\author[Peter Hatfield]{P. W. Hatfield$^{1}$\thanks{peter.hatfield@physics.ox.ac.uk}, S. N. Lindsay$^{1}$, M.J. Jarvis$^{1, 2}$, B.H\"au\ss ler$^{1,3,4}$, M.Vaccari$^{2}$, A.Verma$^{1}$\\
$^{1}$Astrophysics, University of Oxford, Denys Wilkinson Building, Keble Road, Oxford, OX1 3RH, UK\\
$^{2}$Department of Physics, University of the Western Cape, Bellville 7535, South Africa\\
$^{3}$Centre for Astrophysics, Science \& Technology Research Institute, University of Hertfordshire, Hatfield, Herts, AL10 9AB, UK\\
$^{4}$European Southern Observatory, Alonso de Cordova 3107, Vitacura, Casilla 19001, Santiago, Chile\\
}

\begin{document}

\date{In original form 8th May 2015}

\pagerange{\pageref{firstpage}--\pageref{lastpage}} \pubyear{2016}

\maketitle

\label{firstpage}

\begin{abstract}
We present a series of results from a clustering analysis of the first data release of the Visible and Infrared Survey Telescope for Astronomy (VISTA) Deep Extragalactic Observations (VIDEO) survey. VIDEO is the only survey currently capable of probing the bulk of stellar mass in galaxies at redshifts corresponding to the peak of star formation on degree scales. Galaxy clustering is measured with the two-point correlation function, which is calculated using a non parametric kernel based density estimator. We use our measurements to investigate the connection between the galaxies and the host dark matter halo  using a halo occupation distribution methodology, deriving bias, satellite fractions, and typical host halo masses for stellar masses between $10^{9.35}M_{\odot}$ and $10^{10.85}M_{\odot}$, at redshifts $0.5<z<1.7$. Our results show typical halo mass increasing with stellar mass (with moderate scatter) and bias increasing with stellar mass and redshift consistent with previous studies. We find the satellite fraction increased towards low redshifts, from $\sim 5\%$ at $z\sim 1.5$, to  $\sim 20\%$ at $z\sim 0.6$. We combine our results to derive the stellar mass to halo mass ratio for both satellites and centrals over a range of halo masses and find the peak corresponding to the halo mass with maximum star formation efficiency to be $ \sim 2  \times10^{12} M_{\odot}$, finding no evidence for evolution.
 
\end{abstract}


\section{Introduction}

We work in the paradigm of luminous matter (galaxies) being biased tracers of the underlying dark matter distribution.  The growth of cold dark matter (CDM) perturbations is relatively simple to model and understand, both analytically (\citealp{Press1974a}; \citealp{Sheth1999}) and in N-body simulations (\citealp{Warren2006}) as it is thought to be pressure and interaction free. However we cannot observe the dark matter directly; we can only observe the luminous matter following the underlying dark matter distribution in a biased, complex way. Large galaxy surveys allow us to probe this behaviour in a statistical manner, giving insight to the physical processes at play.  Recent wide-field surveys have surveyed the semi-local Universe spectroscopically in great detail e.g. the 2-degree-Field Galaxy Redshift Survey (2dFGRS, \citealp{Peacock2001}), Sloan Digital Sky Survey (SDSS,  \citealp{Zehavi2011}) and the Galaxy And Mass Assembly (GAMA, \citealp{Driver2011}) survey on the kilo-square degree scale, the VIMOS VLT Deep Survey (VVDS, \citealp{LeFevre2013}) and the VIMOS Ultra-Deep Survey (VUDS, \citealp{LeFevre2015}) on degree scales. Similarly, surveys like the United Kingdom Infrared Deep Sky Survey Ultra Deep Survey (UKIDSS-UDS, \citealp{Hartley2013}) and now UltraVISTA (\citealp{McCracken2012}), have probed photometrically very deeply on $\sim 1 \textrm{deg}^2$ scales. The Visible and Infrared Survey Telescope for Astronomy (VISTA) Deep Extragalactic Observations (VIDEO) survey \citep{Jarvis2013} sits fittingly between these two scales of interest as the current leading survey for studying the $z>0.5$ Universe over large scales. It is particularly well suited to investigating many contemporary problems in forming a good all-encompassing model of galaxy evolution. Although modern observational techniques have led to substantial improvements in our understanding of the nature of galaxies and their evolution over cosmic time (e.g. Mo, van den Bosch and White, 2010), there remain many problems in explaining the rich menagerie of galaxies we see in the Universe today. Galaxies come in range of masses spanning several decades (e.g. \citealt{Tomczak2013}), exhibit a range of morphologies (e.g. \citealt{Willett2013}), and can have vastly different star-formation (SF) rates (\citealt{Bergvall2015}). Some exhibit active galactic nucleus (AGN) activity - powerful energetic bursts from accretion onto supermassive black holes, that are thought to impact on the life of the whole galaxy via feedback processes (e.g. \citealt{Fabian2012}). A good model of galaxy evolution must take all these wide ranging phenomena into account (e.g. most semi-analytic and hydrodynamic simulations now incorporate such activity to truncate star formation in massive galaxies, for example \citealp{Dubois2014}) to explain the observations.

VIDEO is particularly well suited to investigating, explaining and constraining many of these problems, as its balance of depth and sky area allows wide scale effects to be probed to earlier times:
\begin{itemize}
  \item It has a multitude of multi-band data for both better constraints on redshift as well extra information like stellar mass and star formation rate of the the galaxies e.g. see \citet{Johnston2015}.
  \item Its depth and high quality photometric redshifts permit the study of galaxies on large scales at $z \sim 1-3$, the peak of star formation in the Universe
  \item Its balance of depth and sky area makes it possible to constrain galaxy behaviour on both sides of the `knee' of the stellar mass function at these crucial redshifts
  \item It has the width and resolution to simultaneously probe the two length-scale regimes of linear and non-linear distributions
  \item It has three separate fields to measure cosmic variance
\end{itemize}

Access to these large-scale effects is crucial for understanding the environment of a galaxy population, which can play an important role in its evolution. Key processes in galaxy evolution are often classified into `nature' and `nurture' effects, e.g. internal processes such as cooling and feedback versus interactions with other galaxies and the local environment - often a variety of processes are needed to explain environmental-based observations such as the morphology-density relation (elliptical galaxies are preferentially found in high-density environments and spiral galaxies in the field; \citealp{Dressler1980}).  A key question is the role of environment and halo mass on quenching (e.g. \citealp{Peng2010}), and how important, or not, processes like strangulation (tidal effects from the gravitational potential allowing the gas in the satellite to leave), ram pressure stripping (removal of gas by `winds' in the hot intra-cluster medium) and harassment (flybys from other galaxies, \citealt{Hirschmann2014}) are. It is also becoming apparent that the larger scale environment, distances well beyond the virial radius of the halo, can have local effects on individual galaxies and lead to large scale correlations, now known as galactic conformity (e.g. \citealt{Weinmann2006}, \citealt{Kauffmann2013} and \citealt{Hearin2015}).

One key probe of the galaxy-dark matter connection is the two-point correlation function (the inverse Fourier transform of the power spectrum) which is a popular measure of the  statistical clustering of galaxies, see \citet{PhillipJamesEdwinP1980}. This is commonly interpreted via the phenomenological model of the halo occupation distribution (HOD, \citealp{COORAY2002a}; \citealp{Zehavi2005}). Typically the galaxy content of a halo is stipulated as some function of the halo mass. Then assuming a halo bias model and halo profile, the correlation function can be predicted, and compared to observations (\citealp{Zheng2005}). Derived parameters from the HOD (minimum mass for galaxy collapse, bias, typical halo mass etc) can then typically be linked to models of galaxy formation and evolution, or compared with results from simulations (e.g. \citealp{Wang2006}). Other probes of the galaxy-DM connection include galaxy-galaxy lensing, which contains information about the host halo profile and can be combined with clustering measurements to great effect (e.g. \citealp{Coupon2015}), and comparison with group catalogues (e.g \citealp{Yang2005}). In this paper we analyse the clustering relations between different galaxy samples to draw out HOD parameters to investigate the galaxy-halo connection to high-$z$ and moderate stellar mass.

The only other survey currently able to probe to similar stellar masses and redshifts on ~ degree scales is UltraVISTA, another public ESO VISTA survey, see \citet{McCracken2012}.  \cite{McCracken2015} perform a clustering analysis in the survey, fitting HOD models, and studying the stellar mass to halo mass ratio. UltraVISTA and the sub-field of VIDEO that we use here probe similar parts of parameter space, giving VIDEO an important role in validating this science on a different field, but in future data releases the surveys will diverge, VIDEO probing wider, and UltraVISTA deeper. Validating clustering measurements on independent fields has particular importance in this instance as the COSMOS field (in which UltraVISTA is carried out) is reported in the literature to have an overabundance of rich structure, and to in general be unrepresentative of similar volumes at the same redshift (e.g \citealt{Meneux2009}, who report a $2-3\sigma$ anomaly by comparison with mock skies). \cite{McCracken2015} explore this complication, speculating that the quasar wall a few degrees away from the field, reported in \citet{Clowes2013}, could give rise to this over-density. They compare clustering measurements in COSMOS with WIRDS data (\citealt{Bielby2014}) over four fields finding agreement on larger scales, but dramatically increased clustering power at small scales in COSMOS at $1<z<1.5$. Not only does this illustrate the importance of having a separate field to confirm these results at these key redshifts over the key epoch when both AGN and SF activity were at their peak, but it also shows that cosmic variance is still a significant factor at these angular scales and that eventually the multiple independent fields of VIDEO are needed. There is also valuable information to be gained by comparing photometry-based results with spectroscopic surveys that have covered the same fields (e.g. VVDS, \citealt{Abbas2010}, and VUDS, \citealt{Durkalec2015a,Durkalec2015}). Spectroscopic surveys have much more accurate redshifts, and can hence get more accurate measurements of clustering, as well as probing effects not present in angular information, in particular redshift space distortions. Conversely, like-for-like spectroscopic surveys typically probe smaller numbers of sources (in a biased manner depending on the selection of the survey sources), ordinarily not probing as deep as an otherwise similar photometric survey. Exploiting the ability of spectroscopic surveys to probe different parts of clustering parameter space in different ways is beneficial for a comprehensive understanding of the galaxy-halo connection and the role of environmental effects at a given epoch.

This paper is organised as follows: first we describe our sample selection from VIDEO (section \ref{sec:OBS}), and discuss how we measure the correlation function (section \ref{sec:selection}). We then discuss our halo occupation model, derived parameters and fitting process in section \ref{sec:hod_description}. We then find the correlation function for a series of sub-samples split by stellar mass, and fit HOD models to these observations. Finally we discuss how derived parameters from the HOD vary with stellar mass and redshift, compare to other studies, and discuss how our measurements will be extended with the full VIDEO survey (section \ref{sec:RESULTS}).

All magnitudes are given in the AB system \cite{Oke1983} and all calculations are in the concordance cosmology $\Omega_{\Lambda}=0.7$, $\Omega_{m}=0.3$ and $H_{0}=70 \text{ km} \text{ s}^{-1} \text{Mpc}^{-1}$ unless otherwise stated.

\section{Observations} \label{sec:OBS}

In this section we describe the optical and near infrared data used to select the galaxies in our sample, and provide information on the photometric redshift and stellar mass estimates that underpin our analysis.

\subsection{VIDEO and CFHTLS}

The VIDEO Survey \citep{Jarvis2013} is one of the 6 public surveys carried out by the VISTA telescope facility in Chile. It covers three fields in the southern hemisphere, each carefully chosen for availability of multiband data, to total 12 deg$^{2}$ when complete. The $5 \sigma$ depths of VIDEO originally planned, and observed to in the XMM3 field, in the five bands are $Z=25.7$, $Y=24.5$, $J=24.4$, $H=24.1$ and $K_s=23.8$ for a 2'' diameter aperture. We note however that the observing plan is now to observe to $Y=25.5$ at the expense of $Z$ due to the inclusion of the fields in the Dark Energy Survey, DES, see \cite{Banerji2015}.

In this study, we use the VIDEO data set combined with data from the T0006 release of the Canada-France-Hawaii Telescope Legacy Survey (CFHTLS) D1 tile \citep{Ilbert2006,Gwyn2012}, which provides photometry with $5 \sigma$ depths of $u^{*}= 27.4$, $g^{\prime}=27.9$, $r^{\prime}=27.6$, $i^{\prime}=27.4$ $z^{\prime}=26.1$ over 1 deg$^{2}$ of the VIDEO XMM3 tile (which will be joined by two other adjacent tiles). Note that the $i^{\prime}$ filter used for CHFTLS is different to the SDSS $i^{\prime}$ filter, and that this data was collected with the \textit{first} MegaCam $i^{\prime}$ filter (during the survey the filter had to be replaced by one with a slightly different response). This data set (and the paramatrisations discussed in section \ref{sec:LePHARE}) has already been used in many extragalactic studies to data (e.g.  \citealp{White2015}; \citealp{Johnston2015}). The infrared VIDEO data for other tiles than XMM3 is now available. However, the publicly available optical data over these fields (CHFTLS Wide-1 and the currently public DES data) are shallower than D1, which would not allow extension to as high redshifts. Future work will extend the analysis in this paper to the wider areas.

\subsection{LePHARE and SExtractor} \label{sec:LePHARE}

The sources in the images are identified using SExtractor, \citep{Bertin1996} source extraction software, with 2'' apertures. See \citet{Jarvis2013} for more details.

The photometric redshifts are calculated using LePHARE \citep{Arnouts1999,Ilbert2006}, which fits spectral energy distribution (SED) templates to the photometry \citep{Jarvis2013}. LePHARE generates a redshift probability density function, stellar masses, star formation rates, CLASS\_STAR (probability of being a star based on compactness) and many other parameters are also calculated. Further information on detection images used, detection thresholds and the construction of the SED templates is given in \citet{Jarvis2013}.

\subsection{Final Sample} \label{sec:final_sample}

SExtractor identifies 481,685 sources in the field with detections in at least  one band. We applied a simple mask to the data in order to cut out areas dominated by foreground stars and any dead pixels. The mask was also applied to the random catalogues used in the calculation of the correlation function (see section \ref{sec:RR}).

Uncertainty in LePHARE parameterisations (photometric redshift estimation etc.) increases at fainter magnitudes, both because the relative error on fluxes is larger for faint objects, and because objects start to be only detected in a few bands. We use a $K$-band cut to remove all galaxies $K_{s}>$23.5. VIDEO has a $90$ percent completeness at this depth \citep{Jarvis2013}.

For removing stars from the sample, SExtractor produces parameter CLASS\_STAR as an indicator of the probability that a given object is a star, based on whether it appears point-like, but this has been shown not to perform well up to the magnitudes we have probed (\citealp{McAlpine2012}, \citealp{White2015}). To eliminate stars from our sample we define a stellar locus as in \citet{Jarvis2013}, following the approach of \citet{Baldry2010},

\begin{equation}
f_{\mathrm{locus}}(x) = \left\{ \begin{array}{l l l} -0.58 & \quad \text{$x<0.4$} \\ -0.58+0.82x-0.21 x^2  & \quad \text{$0.4<x<1.9$}\\ -0.08 & \quad \text{$1.9<x$} \end{array} . \right.\
\end{equation}

We then remove sources with:

\begin{equation}
J-K_{s}<0.12+f_{\mathrm{locus}}(g-i).
\end{equation}

McAlpine et al. estimate this cut leaves stars contributing less than $5\%$ of the sample. The final galaxy sample comprises 97,052 sources after masking, removing stars and making a $K_{s}<$23.5 cut.

\begin{figure}
\includegraphics[scale=0.5]{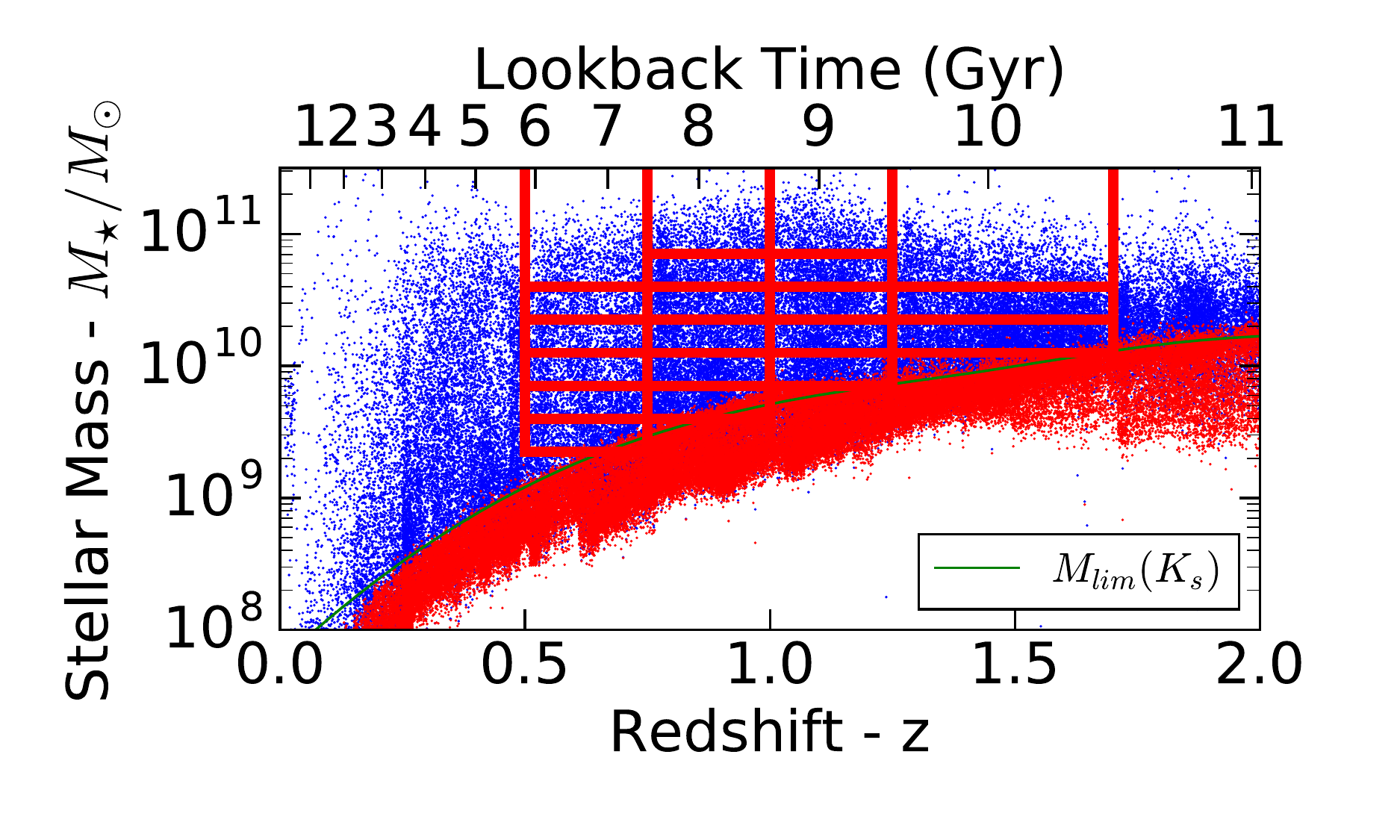}
\caption{The mass and redshift of galaxies, considered after application of the magnitude cut, star exclusion and mask, are shown here in blue. The red points mark the stellar mass limit for all objects that could be detected with our apparent magnitude limit of $K_{s}<23.5$, and the green curve the implied $90\%$ stellar mass completeness limit, following the approach of \protect\cite{Johnston2015}. The red boxes illustrate the redshift and stellar mass selected sub-samples that we consider in subsequent sections}
\label{fig:mass_z}
\end{figure}

\section{The Two-Point Correlation Function} \label{sec:selection}

A range of statistical tools are used to study the interactions between galaxies and to characterise clustering. There exist many ways to measure clustering, in particular nearest neighbour \citep{Bahcall1983}, genus \citep{Park2001} and counts in cells \citep{White1979}. In this study we focus on the two-point correlation function, a measure of how much more likely two galaxies are to be at a given separation than random (in fact counts in cells statistics can be derived from the correlation function). 

The underlying meaningful physical relation is the full three dimensional spatial correlation function; however we only have the observables of angular separations and redshift information. Limber Inversion, (\citealt{Limber:1954zz}) gave an early key way of connecting the two. The two main approaches to connecting the observables to the spatial correlation function are to either calculate the angular correlation function, and compare to angular projections of the model, or to use the redshift information to form the projected correlation function in both transverse and longitudinal directions (which incorporates redshift space distortions, which are normally integrated out), see \citet{Davis1983} and \citet{Fisher1994}. Here we focus on the angular correlation function as the projected correlation function requires very precise knowledge of the redshifts of the sample to avoid being biased, and is in general more appropriate for surveys with more accurate redshifts e.g. spectroscopic surveys.

\subsection{The Angular Correlation Function}\label{sec:ACF_def}

The angular two-point correlation function $\omega(\theta)$ is a measure of how much more likely it is to find two galaxies at a given angular separation than a uniform unclustered Poissonian distribution:

\begin{equation}
dP=\sigma(1+\omega(\theta))d\Omega ,
\end{equation}

where $dP$ is the probability of finding two galaxies at an angular separation $\theta$, $\sigma$ is the surface number density of galaxies, $d\Omega$ is solid angle. We require $\omega(\theta)>-1$ and $\lim_{\theta \to \infty}\omega(\theta) = 0$ for non-negative probabilities and for non-infinite surface densities respectively. 
\subsubsection{Estimating $\omega(\theta)$ Numerically} \label{sec:RR}

The most common way to estimate $\omega(\theta)$ is through calculating $DD(\theta)$, the normalised number of galaxies at a given separation in the real data, and $RR(\theta)$, the corresponding figure for a synthetic catalogue of random galaxies identical to the data catalogue in every way (i.e. occupying the same field) except position. We use the \citet{Landy1993} estimator:

\begin{equation}
\omega(\theta)=\frac{DD-2DR+RR}{RR} ,
\end{equation}

which also uses $DR(\theta)$,  data to random pairs, as it has a lower variance (as an estimator) and takes better account of edge effects, although there are other estimators (as discussed and compared in \citealt{Kerscher2000}).

By averaging over multiple average data sets and using $\overline{RR(\theta)}$ or by letting the number of random data points go to infinity the error in $RR(\theta)$ can be considered zero e.g. essentially becomes a function of the field geometry. We use 500,000 random data points in this study. This leaves $DD(\theta)$ as the main source of variance in our estimation, and is often given as the Poisson error in the DD counts:

\begin{equation}
\Delta \omega=\frac{1+\omega(\theta)}{\sqrt{DD}} .
\end{equation}

However this naive approach can significantly underestimate the uncertainty because adjacent $DD$ bins are correlated. More rigorous approaches therefore rely on bootstrap methods. The `jack-knife' method consists of blocking off segments of the field and recalculating to see how much the estimate of the correlation functions changes. `Bootstrap resampling' consists of sampling the galaxies with replacement from the dataset and recalculating, see \citet{Ling1986}. Repetition of this process allows the variance of the $\omega(\theta)$ values to be estimated. \citet{Lindsay2014} found Poisson errors were a factor of 1.5 to 2 smaller than those estimated with bootstrap. In this paper we use 100 bootstrap resamplings to estimate the uncertainty at the 16th and 84th percentiles of the resampling.

The finite size of the survey area results in a negative offset to the true correlation function, known as the integral constraint:

\begin{equation}
\omega_{obs}(\theta)=\omega_{true}(\theta)-K_{IC} .
\end{equation}

$K_{IC}$ has an analytic expression from \citet{Groth1977}

\begin{equation}
K_{IC}=\frac{1}{\Omega^{2}}\iint\omega_{true}(\theta)d\Omega_{1}d\Omega_2 ,
\end{equation}

where $d\Omega_{1}d\Omega_2$ denotes integrating twice over the field solid angle, which can be estimated numerically \citep{Roche1999} by:

\begin{equation}
K_{IC}=\frac{\Sigma RR(\theta)\omega_{true}(\theta)}{\Sigma RR(\theta)} .
\label{eq:IC_expression}
\end{equation}

The integral constraint has the effect of reducing the measured correlation function at large angles and steepening the gradient. Therefore, when fitting the correlation functions, we treat the constraint as part of the model and fit data to the theoretical observed function, as in \citet{Beutler2011}.

\subsection{Non-Parametric Estimation}\label{sec:non_param}

Approaches to calculating the correlation function conventionally involve binning; the galaxy angular separations are put into angular distance bins (often spaced logarithmically). Although advantageous in terms of simplicity to calculate, and clearer interpretation, binning data is non-ideal in the sense that it i) loses information and ii) can involve arbitrary choice of bin size. Here we present an alternative estimator that finds the correlation function as a continuous function.

The approach we used was to implement a non-parametric method for the estimation of $DD(\theta)$, $DR(\theta)$ and $RR(\theta)$, and then use the estimator of Landy and Szalay as per usual. We use here the kernel based density estimator of Parzen and Rosenblatt \citep{Parzen1962,Rosenblatt1956} on the set of angular separations to find $DD$, $DR$ and $RR$ separately, and then choose kernel bandwidth to minimise the mean integrated squared error (MISE) for each. Heuristically the process can be described thus: first the $\frac{1}{2} N^2$ galaxy separations are calculated. However rather than being binned by angular separation, the distribution is calculated by summing kernel distributions (e.g. normal, top hat or tricube etc.) centred on each point, and kernel width replaces the role of bin size. If the width of the kernel is too large, the data are over smoothed, and features are lost. If the width is too small, the data is too noisy. There exists an optimal choice that minimises the expected error on this method as an estimator of the true distribution.

We give a brief description of how to choose optimal smoothing parameters as described in \cite{Parzen1962}. Suppose $f(x)$ is the true function that we are attempting to estimate (in our context it could be $DD(\theta)$) and that $\hat{f}(x)$ is our estimation of the function from the data. The quantity to be minimised is the expected error accumulated over all x, the MISE:

\begin{equation}
\mathrm{MISE}=\mathbb{E} \left(\int (f(x)-\hat{f}(x))^2 \mathrm{d}x \right)  ,
\end{equation}

which can be re-arranged to:

\begin{equation}
\mathrm{MISE}=\int b^{2}(x) \mathrm{d}x + \int v(x) \mathrm{d}x ,
\end{equation}

where

\begin{equation}
b(x)=\mathbb{E} (\hat{f}(x))-f(x) ,
\end{equation}

the bias of the estimator at a point and

\begin{equation}
v(x)=\mathbb{V}(\hat{f}(x)) ,
\end{equation}

the variance of the estimator at a point. Hence MISE is a function of the data and the smoothing parameter $h$. Minimising MISE is a compromise between minimising bias and minimising estimator variance.

If our data are $X_i$ (in our case galaxy separations) and our kernel is K (a smooth symmetric function around zero that integrates to 1 and goes to zero sufficiently fast; e.g. we use a Gaussian), then our estimate of the function becomes:

\begin{equation}
\hat{f}(x) = \frac{1}{nh} \sum_{i} K\left( \frac{x-X_i}{h} \right) ,
\end{equation}

where $h$ can be chosen to be the standard deviation of the kernel. \cite{Parzen1962} show that this estimate is consistent and that the MISE goes as:

\begin{equation}
\mathrm{MISE} \approx \frac{1}{4} c^{2}_{1} h^{4} \int(f''(x))^2 \mathrm{d}x + \frac{\int K^{2} (x) \mathrm{d}x }{nh} ,
\end{equation}

and the optimal $h$ (by differentiation) is

\begin{equation}
h_{*} = \left( \frac{c_2}{c_{1}^{2} A(f) n} \right) ^{1/5} ,
\end{equation}

where $c_{1}=\int x^{2} K(x) \mathrm{d}x$, $c_{2}=\int K(x)^2 \mathrm{d}x$ and $A(f)=\int (f''(x))^2 \mathrm{d}x$.

Calculating the optimal $h$ value in this manner formally depends on knowing the true distribution e.g. if $f$ oscillates wildly with high frequency $A(f)$ is high, necessitating a smaller $h$ to pick out features). We calculate this heuristically by doing a first run with a trial value of $h$, fitting a power law to the resulting correlation function, and using this as the true distribution for the purposes of finding $h$. We then subsequently use the estimate of the correlation function that results as our estimate of the true value. We find typical suitable values of $h$ are of order 0.1-0.2 dex in angular separation, of comparable order to bin sizes most authors choose heuristically.

There do exist entirely data driven cross-validation techniques of choosing $h$ optimally that we do not discuss here; see \citet{Bowman1984} for a discussion. For continuous estimation of error (which now takes the form of a band along our estimation of the function) we repeat the discussed process on bootstrapped data sets and take the 16th and 84th percentiles point-wise of our multiple estimations of the correlation function.

\begin{figure}
\includegraphics[scale=0.5]{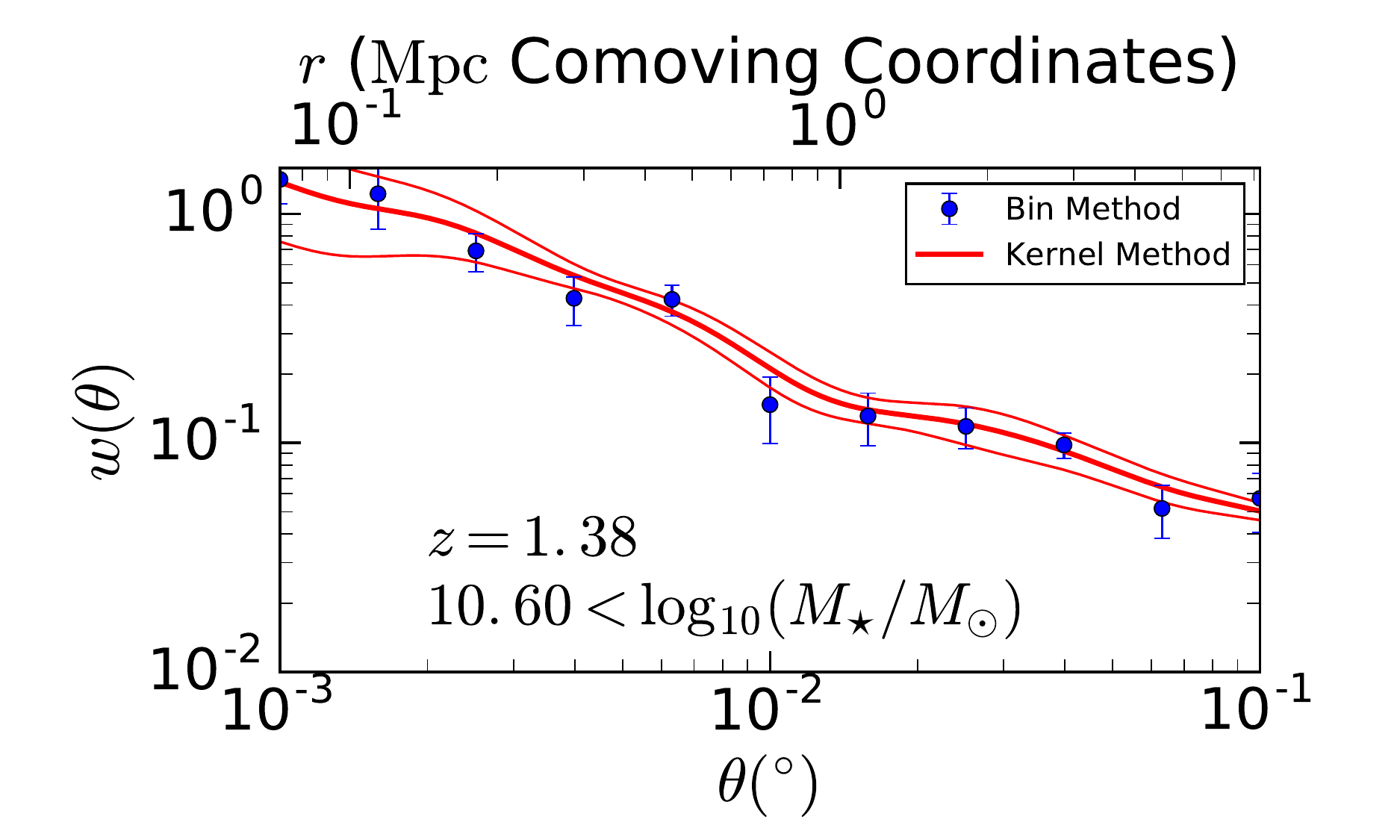}
\caption{Illustration of the agreement of the binning approach and the non-parametric approach to correlation function calculation for a sample with $1.25<z<1.5$, $ M_{\star} >10^{10.6} M_{\odot} $. The error bars (the secondary lines in the case of the kernel method) for both the discrete and the continuous methods represent the 16th and 84th percentiles from bootstrapping}
\label{NONPARAM}
\end{figure}

To confirm that we are consistent with the binning method, the correlation function was calculated using the binning approach as well as the kernel approach for a sample with $1.25<z<1.5$, $ M_{\star} > 10^{10.6} M_{\odot} $ sample, see fig \ref{NONPARAM}. A way of viewing the kernel approach is that it is essentially the same as binning - except binning uses a top hat kernel of arbitrary size for each data point, and does not always place the kernel directly on top of the data.

\subsection{Redshift Probability Density Distributions} \label{sec:redshift_dist}

The correlation function is often calculated assuming just the best redshift value for the source, without consideration of the uncertainty in the measurement e.g for sources with broad redshift probability density functions (pdfs) the chance of the object being in a different redshift bin to its best fit is not accounted for. We take the approach of \cite{Arnouts2002} and for each redshift bin we assign galaxies a weight corresponding to the probability of the galaxy being in that redshift range according the the LePhare redshift pdfs, e.g.

\begin{equation} \label{eq:weights}
W_{i}=\int_{z_{lower}}^{z_{upper}} p_{i}(z) \,\mathrm{d}z ,
\end{equation}

and

\begin{equation}
DD(\theta)=\frac{2}{n(n-1)h}\sum_{i,j} W_{i} W_{j} K\left(\frac{\theta - d(G_i,G_j)}{h}\right) ,
\end{equation}
where $K$ is the chosen kernel, $h$ is the kernel width, $d(G_i,G_j)$ is the angular separation of galaxies $i$ and $j$ and $n$ is the number of galaxies, $n(n-1)/2$ being the number of galaxy pairs. In the limit of highly accurate redshifts, this method reduces to the approach of just working with galaxies where the probability density function has its peak in the bin. If we replace the probability density function with a Dirac delta function at the peak, as would be the case in a spectroscopic survey, the approaches coincide.

\section{Halo Occupation Distribution Modelling} \label{sec:hod_description}

Halo Occupation Distribution (HOD) descriptions of correlation functions have seen great success in recent years in modelling the correlation function to high degrees of precision, and giving physical results in agreement with other methods (e.g. \citealp{Simon2008}, \citealp{Coupon2015}). Models typically prescribe the mean number of galaxies in a halo as a function of halo mass, assume the occupation number has a Poissonian distribution, and assume that these galaxies trace the halo profile. Then the HOD model, choice of halo profile, halo mass function and a bias proscription can be translated into a spatial correlation function, and then projected to an angular correlation function. Parametrising the HOD allows physical information to be extracted via some fitting process. Variants include fitting simultaneously with cosmology \citep{Bosch2012}, varying the compactness of the profile the galaxies follow, allowing the occupation statistics to be non-Poissonian and investigating if different galaxy samples occupy the halos independently \citep{Simon2008}. It is also possible to fit HOD models using galaxy-galaxy weak lensing data (e.g. \citealp{Coupon2015}), background counts in halo catalogues (e.g. \citealp{Rodriguez2015}), and even abundance matching techniques (e.g. \citealp{Guo2015}). Our model and approach follows closely that in \citet{Coupon2012} and \citet{McCracken2015}.

\subsection{The Model} \label{sec:LIMBER}

We use the 5 parameter model of \cite{Zheng2005}, assuming a Navaro-Frenk-White  profile \citep*{Navarro1997} and a \citet{Tinker2010} bias model. We use the \textit{halomod} python package\footnote{https://github.com/steven-murray/halomod} to calculate correlation functions.The five parameters are;

\begin{itemize}
  \item $M_{\mathrm{min}}$, minimum halo mass required for the halo to host a central galaxy
  \item $M_{1}$ the typical halo mass for satellites to start forming
  \item $\alpha$ as the power law index for how the number of satellites grows with the halo mass
  \item $\sigma_{\log_{10}M}$ parametrises how discrete the cut off in halo mass for forming a central galaxy is, and
  \item $M_{0}$ a halo mass below which no satellites are formed.
\end{itemize}

The number of central and satellite galaxies, as well as total number, are parametrised by the following equations:

\begin{equation}
\langle N_{\mathrm{central}} \rangle = \frac{1}{2} \left(1+\textrm{erf} \left(  \frac{ \log_{10} M_{\mathrm{\mathrm{halo}}} - \log_{10} M_{\mathrm{min}} }{ \sigma_{ \log_{10} M} } \right) \right)
\end{equation}
\begin{equation}
\langle N_{\mathrm{sat}} \rangle = \langle N_{\mathrm{central}} \rangle \times \left(\frac{M_{\mathrm{\mathrm{halo}}}-M_{0}}{M_{1}}\right)^{\alpha}
\end{equation}
\begin{equation}
\langle N_{\mathrm{total}} \rangle = \langle N_{\mathrm{central}} \rangle + \langle N_{\mathrm{sat}} \rangle .
\end{equation}

Thus the number of central galaxies as a function of halo mass behaves as a softened step function and the number of satellites is a power law that initiates at a characteristic mass. The equations only allow there to be satellite galaxies when there is a central galaxy.

Given a set of parameters, the model correlation function is constructed from a 1-halo term on small scales describing non-linear clustering within a halo constructed from the halo profile, and a 2-halo term on large scales describing clustering between halos, constructed from the bias prescription and dark matter power spectrum. The transition between the two regimes is typically at approximately $1$Mpc, or around $0.05^{\circ}$ in angular space at these redshifts. Within a halo, the first galaxy is assumed to be at the centre of the halo (the central), and the positions of all subsequent galaxies (the satellites) trace the profile of the halo. The 1-halo term can thus be further broken down into a central-satellite term, formed by convolving a NFW profile with a point and weighted by $\langle N_{\mathrm{sat}} \rangle$, and a satellite-satellite term, formed by convolving a NFW profile with itself and weighted by $\langle N_{\mathrm{sat}}(N_{\mathrm{sat}}-1) \rangle$. The net effect of this is to add power at smaller radii. This expression for a single halo is then averaged for all halo masses by integrating, weighting by the halo mass function. The 2-halo term is constructed by finding the inverse Fourier transform of the dark matter power spectrum multiplied by the square of the `averaged' bias. The averaged bias is found by multiplying the number of galaxies in a given halo mass by the bias of that halo, and then averaging by multiplying by the halo mass function and integrating over all halo masses to average. The 1-halo and 2-halo terms are then summed to find the spatial correlation function, and then projected using the redshift distribution to form the angular correlation function. See \cite{Coupon2012} for a more in depth description of this process. The occupation numbers are assumed to be Poissonian when calculating variables like $\langle N_{\mathrm{sat}}(N_{\mathrm{sat}}-1) \rangle$ etc.  

\subsection{Incorporating Stellar Mass Ranges in the Model} \label{sec:LIMBER}

Angular correlation functions, and hence the derived HOD models, are highly dependent on the magnitude cut or stellar mass range of the galaxies, which is to be expected as different galaxy samples typically exist in different halos. The approach we use to build up a self consistent picture of how galaxies of different stellar masses occupy the halos is to calculate the correlation function for all the galaxies above a certain mass threshold, for a range of thresholds. We then expect the HOD models to be consistent e.g. a sample of a higher stellar mass threshold does not predict more galaxies at a given halo mass than a lower stellar mass threshold! An alternate approach would be to calculate the correlation function for stellar mass ranges as in \citet{Coupon2015}, which reduces the covariance between measurements. This, however, is better suited to fitting a global occupation model where the halo occupation is a conditional function of the stellar mass given the halo mass, because otherwise the occupation number as a function of halo mass is not straightforward when there is an upper bound of stellar mass.

\subsection{Derived Parameters} \label{sec:derived}

A halo occupation model also gives the galaxy bias and typical host halo mass. Galaxy bias describes how over-dense or under-dense galaxies are compared to dark matter and can be found by comparing the galaxy and dark matter correlation functions:

\begin{equation}
b=\frac{\delta_{\mathrm{g}}}{\delta_{\mathrm{DM}}}=\sqrt{\frac{\xi_{\mathrm{g}}}{\xi_{\mathrm{DM}}}} ,
\end{equation}

where $b$ is the galaxy bias, $\delta_g$ is the local galaxy over-density, $\delta_{\mathrm{DM}}$ is the local dark matter over-density, $\xi_{\mathrm{g}}$ is the galaxy spatial correlation function and $\xi_{\mathrm{g}}$ the dark matter spatial correlation function. It is scale dependent, but settles to a constant value at high separations in the linear regime for standard cosmological models (e.g. \citealp{McCracken2015}). A bias at a given redshift also corresponds to a typical halo mass; both the bias and typical halo mass are derived quantities from the HOD model (see \citealp{Zehavi2005}).

We also calculate for completeness $r_{0}$, the comoving separation at which the spatial correlation function (for the best fit parameters) is unity. This is useful as it operates as a monotone one-dimensional measurement of clustering (as opposed to HOD parameters, which cannot be summarised with one number). It also allows comparison with studies that study correlation functions with a power law, which normally model the spatial correlation function as $\xi(r)=(r/r_{0})^{-\gamma}$. Note $r_{0}$ in the context of this paper is a derived parameter of the HOD model; it does not come from a power law fit to the correlation function.

\subsection{Projecting from $3D$ and Choice of $N(z)$} \label{sec:LIMBER}

\begin{figure}
\includegraphics[scale=0.5]{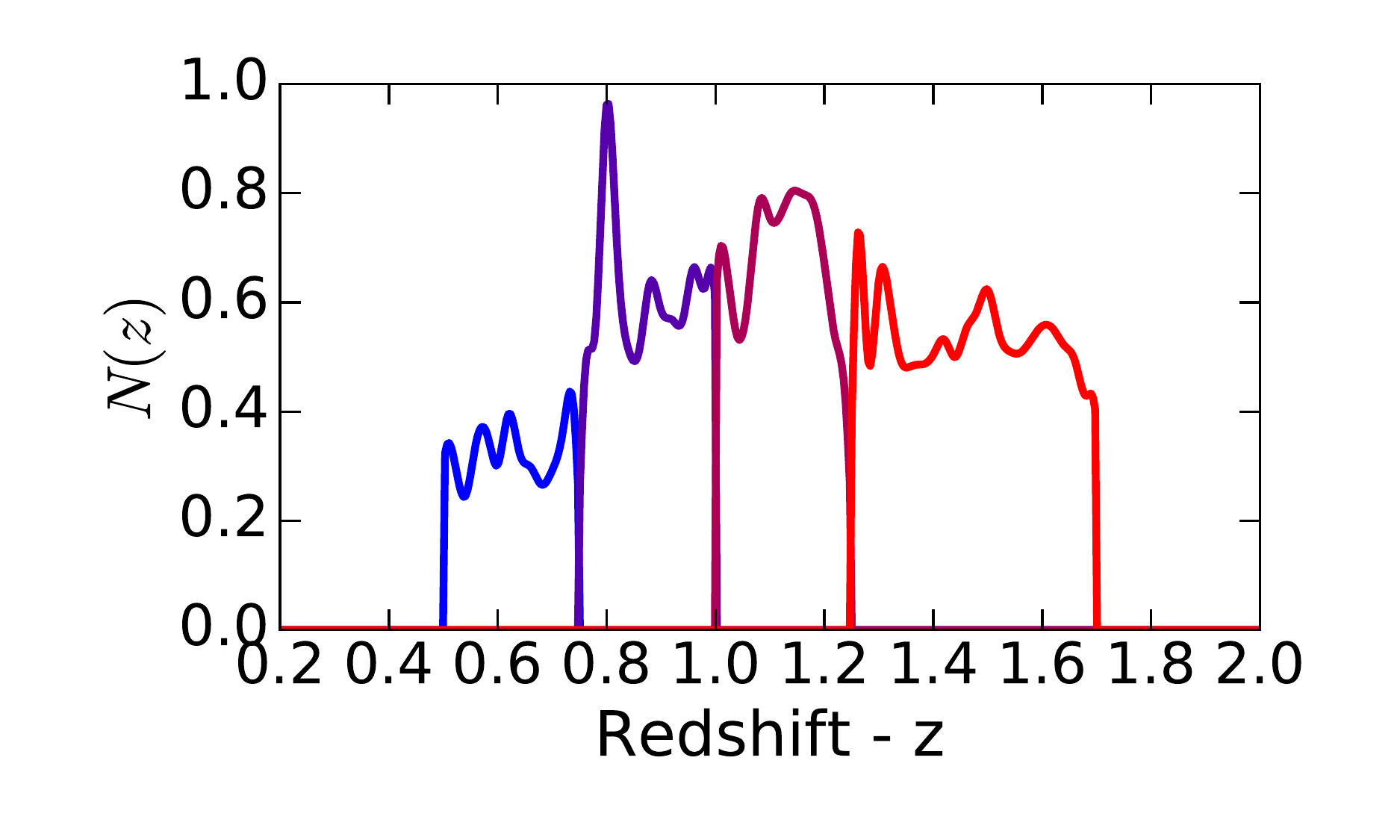}
\caption{The redshift distributions used in our analysis for each redshift bin (arbitrary normalisation).}
\label{fig:redshift_distributions}
\end{figure}

Projecting the spatial correlation function to angular space requires input of $N(z)$, the redshift distribution of the galaxies in the sample. If the redshift is known precisely for each galaxy then this is unambiguous. In \citet{Lindsay2014}, for each redshift bin, the sum of the pdfs that have their peak in that bin is used e.g. there is some contribution from outside the bin. However, with the system of weights, we have only used the part of the probability density that is in each individual bin. Therefore we use the sum of the pdfs just in the part of parameter space considered marginalised over all other variables, which leads to sharp cutoffs, see fig \ref{fig:redshift_distributions}.

We note a sharp peak at $z \sim 0.8$, which could indicate the presence of a large structure at this redshift.

\subsection{MCMC Fitting Process} \label{sec:MCMC}
We use emcee\footnote{http://dan.iel.fm/emcee/current/} (\citealp{Foreman-Mackey2012}) to provide a Markov chain Monte Carlo sampling the parameter space to fit our correlation functions. We use a uniform prior over $0.5<\alpha<2.5$, $0<\sigma<0.6$, $10<\log_{10}{(M_{\mathrm{min}}/M_{\odot})}<15$, $\log_{10}{(M_{\mathrm{min}}/M_{\odot})}<\log_{10}{(M_{1}/M_{\odot})}<17$ and $8<\log_{10}{(M_{0}/M_{\odot})}<\log_{10}{(M_{1}/M_{\odot})}$ (uniform in log space for mass). We used 20 walkers with 1000 steps, which have starting positions drawn uniformly from the prior.

Our likelihood is calculated using $\chi^{2}$ from both the correlation function and the galaxy abundance,

\begin{equation}  \label{eq:chi}
\chi^{2}= \frac{[n_{\mathrm{\mathrm{gal}}}^{\mathrm{obs}}-n_{\mathrm{\mathrm{gal}}}^{\mathrm{model}}]^{2}}{\sigma_{n}^{2}}  +  \sum\limits_{i} \frac{[\omega^{\mathrm{obs}}(\theta_{i})-\omega^{\mathrm{model}}(\theta_{i})]^{2}}{\sigma_{w_{i}}^{2}} ,
\end{equation}

 where $n_{\mathrm{\mathrm{gal}}}^{\mathrm{obs}}$ is the observed number of galaxies in the sample, $n_{\mathrm{\mathrm{gal}}}^{\mathrm{model}}$ is the predicted number of galaxies in that redshift range for a given model, $\sigma_{n}$ is the error on the number counts including both Poisson noise and cosmic variance, $\theta_{i}$ are the angular scales we fit at, $\omega^{\mathrm{obs}}$ is the observed angular correlation function, $\omega^{\mathrm{model}}$ is the angular correlation function of a given model, and $\sigma_{w_{i}}$ is the error on the measured correlation function from the bootstrapping.

As we estimate $\omega(\theta)$ as a continuous function, covariance between measurements is less straightforward. We work around this by fitting to points equally separated in log-space between $0.001\mathrm{ ^{\circ}}$ and $0.1\mathrm{ ^{\circ}}$, with the separation chosen to be greater than the smoothing scale of the non-parametric estimation to minimise covariance between points. We calculate the error on the number counts (which must include both Poisson noise and cosmic variance) as in \citet{Trenti2008}.

Finding $n_{\mathrm{\mathrm{gal}}}^{\mathrm{obs}}$ is complicated by the mask used to remove defects in the field (discussed section \ref{sec:final_sample}), as well as the fact that each galaxy is effectively in multiple redshift bins (discussed section \ref{sec:redshift_dist}). We account for this by instead of counting the galaxies, counting the weights, and then rescaling by the amount of field lost by the mask:

\begin{equation}  \label{eq:ng}
n_{\mathrm{\mathrm{gal}}}^{\mathrm{obs}}=  \frac{\sum\limits_{i} W_{i}}{1-A},
\end{equation}

where $A$ is the fraction of the field covered by the mask (0.03 in our case) and $W_{i}$ are the weights from equation \ref{eq:weights}.

\section{Results} \label{sec:RESULTS}

We show our results for the halo occupation modelling in table 1.

\begin{landscape}
\begin{table}
\caption {Parameters from our angular correlation function evaluation and HOD fitting, with corresponding $\chi^2/\textrm{d.o.f.}$ values for the fits} \label{TABLE2}
\begin{tabular}{@{}cccccccccccc@{}}
\toprule
Stellar Mass Threshold	& $n_{g}$	& $M_{min}$	& $M_{1}$	& $\alpha $	& $\sigma$  & $M_{0}$ & $b$	& $f_{s}$  & {$r_{0}$} & $\chi^2/\textrm{d.o.f.}$ \\
\midrule
{$0.50<z<0.75$} & {$z_{med}=0.62$}\\\midrule
{$9.25$} & {6535} & {$11.7\substack{+0.063 \\ -0.075}$} & {$12.9\substack{+0.13 \\ -0.21}$} & {$0.948\substack{+0.1 \\ -0.16}$} & {$0.5\substack{+0.078 \\ -0.2}$} & {$11.8\substack{+0.46 \\ -1.3}$} & {$1.17\substack{+0.018 \\ -0.012}$} & {$0.213\substack{+0.044 \\ -0.019}$} & {$6.21\substack{+0.12 \\ -0.085}$} & 1.33\\
{$9.5$} & {5061} & {$11.8\substack{+0.09 \\ -0.083}$} & {$12.8\substack{+0.22 \\ -0.16}$} & {$0.82\substack{+0.15 \\ -0.1}$} & {$0.412\substack{+0.14 \\ -0.23}$} & {$12.3\substack{+0.19 \\ -0.34}$} & {$1.22\substack{+0.024 \\ -0.017}$} & {$0.196\substack{+0.014 \\ -0.016}$} & {$6.56\substack{+0.17 \\ -0.12}$} & 0.661\\
{$9.75$} & {3877} & {$12.0\substack{+0.071 \\ -0.077}$} & {$13.1\substack{+0.16 \\ -0.22}$} & {$0.948\substack{+0.14 \\ -0.19}$} & {$0.455\substack{+0.11 \\ -0.17}$} & {$12.1\substack{+0.41 \\ -1.4}$} & {$1.24\substack{+0.022 \\ -0.016}$} & {$0.191\substack{+0.033 \\ -0.018}$} & {$6.71\substack{+0.14 \\ -0.11}$} & 0.918\\
{$10$} & {2847} & {$12.1\substack{+0.053 \\ -0.059}$} & {$13.1\substack{+0.15 \\ -0.23}$} & {$0.887\substack{+0.16 \\ -0.21}$} & {$0.539\substack{+0.047 \\ -0.13}$} & {$12.2\substack{+0.4 \\ -1.5}$} & {$1.26\substack{+0.016 \\ -0.013}$} & {$0.193\substack{+0.061 \\ -0.021}$} & {$6.92\substack{+0.12 \\ -0.098}$} & 1.66\\
{$10.25$} & {1884} & {$12.3\substack{+0.061 \\ -0.084}$} & {$13.4\substack{+0.11 \\ -0.2}$} & {$0.996\substack{+0.11 \\ -0.3}$} & {$0.507\substack{+0.072 \\ -0.18}$} & {$11.9\substack{+0.76 \\ -2.2}$} & {$1.33\substack{+0.024 \\ -0.019}$} & {$0.175\substack{+0.073 \\ -0.026}$} & {$7.38\substack{+0.19 \\ -0.14}$} & 1.27\\
{$10.5$} & {1022} & {$12.5\substack{+0.088 \\ -0.09}$} & {$13.7\substack{+0.087 \\ -0.19}$} & {$1.23\substack{+0.14 \\ -0.35}$} & {$0.418\substack{+0.14 \\ -0.23}$} & {$12.0\substack{+0.95 \\ -2.5}$} & {$1.47\substack{+0.036 \\ -0.034}$} & {$0.129\substack{+0.026 \\ -0.018}$} & {$8.52\substack{+0.27 \\ -0.26}$} & 0.839\\
\midrule
{$0.75<z<1.00$} & {$z_{med}=0.88$}\\\midrule
{$9.5$} & {9791} & {$11.7\substack{+0.052 \\ -0.08}$} & {$12.9\substack{+0.17 \\ -0.24}$} & {$0.942\substack{+0.16 \\ -0.19}$} & {$0.516\substack{+0.065 \\ -0.21}$} & {$11.9\substack{+0.42 \\ -1.4}$} & {$1.24\substack{+0.022 \\ -0.013}$} & {$0.155\substack{+0.028 \\ -0.015}$} & {$5.68\substack{+0.13 \\ -0.076}$} & 0.862\\
{$9.75$} & {7365} & {$11.8\substack{+0.049 \\ -0.058}$} & {$12.9\substack{+0.27 \\ -0.22}$} & {$0.796\substack{+0.27 \\ -0.16}$} & {$0.529\substack{+0.055 \\ -0.12}$} & {$12.3\substack{+0.21 \\ -0.83}$} & {$1.27\substack{+0.017 \\ -0.013}$} & {$0.149\substack{+0.015 \\ -0.014}$} & {$5.89\substack{+0.1 \\ -0.073}$} & 0.896\\
{$10$} & {5453} & {$12.0\substack{+0.049 \\ -0.069}$} & {$13.0\substack{+0.2 \\ -0.22}$} & {$0.806\substack{+0.24 \\ -0.19}$} & {$0.523\substack{+0.059 \\ -0.18}$} & {$12.3\substack{+0.26 \\ -0.75}$} & {$1.32\substack{+0.022 \\ -0.017}$} & {$0.144\substack{+0.017 \\ -0.013}$} & {$6.17\substack{+0.14 \\ -0.1}$} & 1.2\\
{$10.25$} & {3824} & {$12.1\substack{+0.049 \\ -0.061}$} & {$13.4\substack{+0.11 \\ -0.19}$} & {$0.92\substack{+0.19 \\ -0.29}$} & {$0.53\substack{+0.052 \\ -0.14}$} & {$12.1\substack{+0.43 \\ -2.1}$} & {$1.35\substack{+0.021 \\ -0.015}$} & {$0.119\substack{+0.043 \\ -0.013}$} & {$6.35\substack{+0.13 \\ -0.093}$} & 1.43\\
{$10.5$} & {2330} & {$12.3\substack{+0.074 \\ -0.082}$} & {$13.7\substack{+0.14 \\ -0.16}$} & {$0.965\substack{+0.21 \\ -0.32}$} & {$0.426\substack{+0.12 \\ -0.18}$} & {$12.3\substack{+0.42 \\ -1.9}$} & {$1.46\substack{+0.029 \\ -0.025}$} & {$0.0945\substack{+0.026 \\ -0.013}$} & {$7.11\substack{+0.19 \\ -0.16}$} & 0.549\\
{$10.75$} & {1023} & {$12.6\substack{+0.03 \\ -0.038}$} & {$14.2\substack{+0.22 \\ -0.18}$} & {$0.633\substack{+0.21 \\ -0.11}$} & {$0.585\substack{+0.012 \\ -0.036}$} & {$12.3\substack{+0.27 \\ -1.7}$} & {$1.52\substack{+0.017 \\ -0.015}$} & {$0.101\substack{+0.12 \\ -0.018}$} & {$7.52\substack{+0.11 \\ -0.1}$} & 4.33\\
\midrule
{$1.00<z<1.25$} & {$z_{med}=1.12$}\\\midrule
{$9.75$} & {7512} & {$11.8\substack{+0.11 \\ -0.084}$} & {$13.2\substack{+0.15 \\ -0.074}$} & {$1.21\substack{+0.074 \\ -0.14}$} & {$0.339\substack{+0.17 \\ -0.2}$} & {$10.4\substack{+1.5 \\ -1.4}$} & {$1.46\substack{+0.026 \\ -0.027}$} & {$0.13\substack{+0.029 \\ -0.022}$} & {$6.09\substack{+0.14 \\ -0.15}$} & 0.229\\
{$10$} & {5529} & {$12.0\substack{+0.063 \\ -0.082}$} & {$13.2\substack{+0.14 \\ -0.24}$} & {$1.04\substack{+0.22 \\ -0.29}$} & {$0.467\substack{+0.097 \\ -0.2}$} & {$12.2\substack{+0.42 \\ -1.1}$} & {$1.5\substack{+0.029 \\ -0.021}$} & {$0.115\substack{+0.018 \\ -0.016}$} & {$6.34\substack{+0.17 \\ -0.11}$} & 0.521\\
{$10.25$} & {3892} & {$12.2\substack{+0.051 \\ -0.062}$} & {$13.4\substack{+0.071 \\ -0.06}$} & {$1.15\substack{+0.082 \\ -0.19}$} & {$0.535\substack{+0.049 \\ -0.11}$} & {$11.1\substack{+1.1 \\ -1.8}$} & {$1.53\substack{+0.021 \\ -0.018}$} & {$0.115\substack{+0.037 \\ -0.023}$} & {$6.48\substack{+0.12 \\ -0.11}$} & 0.689\\
{$10.5$} & {2412} & {$12.3\substack{+0.05 \\ -0.089}$} & {$13.7\substack{+0.13 \\ -0.074}$} & {$1.07\substack{+0.13 \\ -0.17}$} & {$0.532\substack{+0.052 \\ -0.18}$} & {$10.7\substack{+1.3 \\ -1.7}$} & {$1.62\substack{+0.036 \\ -0.023}$} & {$0.13\substack{+0.066 \\ -0.042}$} & {$7.06\substack{+0.21 \\ -0.14}$} & 0.937\\
{$10.75$} & {1064} & {$12.6\substack{+0.091 \\ -0.091}$} & {$13.9\substack{+0.29 \\ -0.11}$} & {$1.25\substack{+0.28 \\ -0.5}$} & {$0.378\substack{+0.17 \\ -0.21}$} & {$11.8\substack{+1.1 \\ -2.1}$} & {$1.93\substack{+0.058 \\ -0.054}$} & {$0.068\substack{+0.021 \\ -0.012}$} & {$8.95\substack{+0.36 \\ -0.34}$} & 0.635\\
\midrule
{$1.25<z<1.70$} & {$z_{med}=1.48$}\\\midrule
{$10$} & {10800} & {$11.9\substack{+0.065 \\ -0.036}$} & {$13.3\substack{+0.057 \\ -0.12}$} & {$1.36\substack{+0.11 \\ -0.32}$} & {$0.188\substack{+0.15 \\ -0.13}$} & {$11.4\substack{+0.98 \\ -2.0}$} & {$1.78\substack{+0.021 \\ -0.027}$} & {$0.076\substack{+0.011 \\ -0.0087}$} & {$6.47\substack{+0.11 \\ -0.14}$} & 0.437\\
{$10.25$} & {5875} & {$12.1\substack{+0.053 \\ -0.033}$} & {$13.5\substack{+0.075 \\ -0.16}$} & {$1.41\substack{+0.17 \\ -0.46}$} & {$0.119\substack{+0.14 \\ -0.085}$} & {$11.8\substack{+0.88 \\ -2.4}$} & {$1.98\substack{+0.027 \\ -0.026}$} & {$0.0575\substack{+0.0081 \\ -0.0076}$} & {$7.54\substack{+0.14 \\ -0.14}$} & 0.932\\
{$10.5$} & {2542} & {$12.4\substack{+0.088 \\ -0.068}$} & {$14.0\substack{+0.45 \\ -0.16}$} & {$1.06\substack{+0.38 \\ -0.43}$} & {$0.329\substack{+0.13 \\ -0.15}$} & {$12.2\substack{+0.67 \\ -2.2}$} & {$2.16\substack{+0.036 \\ -0.044}$} & {$0.0369\substack{+0.016 \\ -0.0076}$} & {$8.52\substack{+0.2 \\ -0.24}$} & 1.41\\
\bottomrule
\end{tabular}
\end{table}
\end{landscape}

\subsection{Redshift and Stellar Mass Selection} \label{sec:RESULTS_measurements}

\begin{figure*}
\includegraphics[scale=0.45]{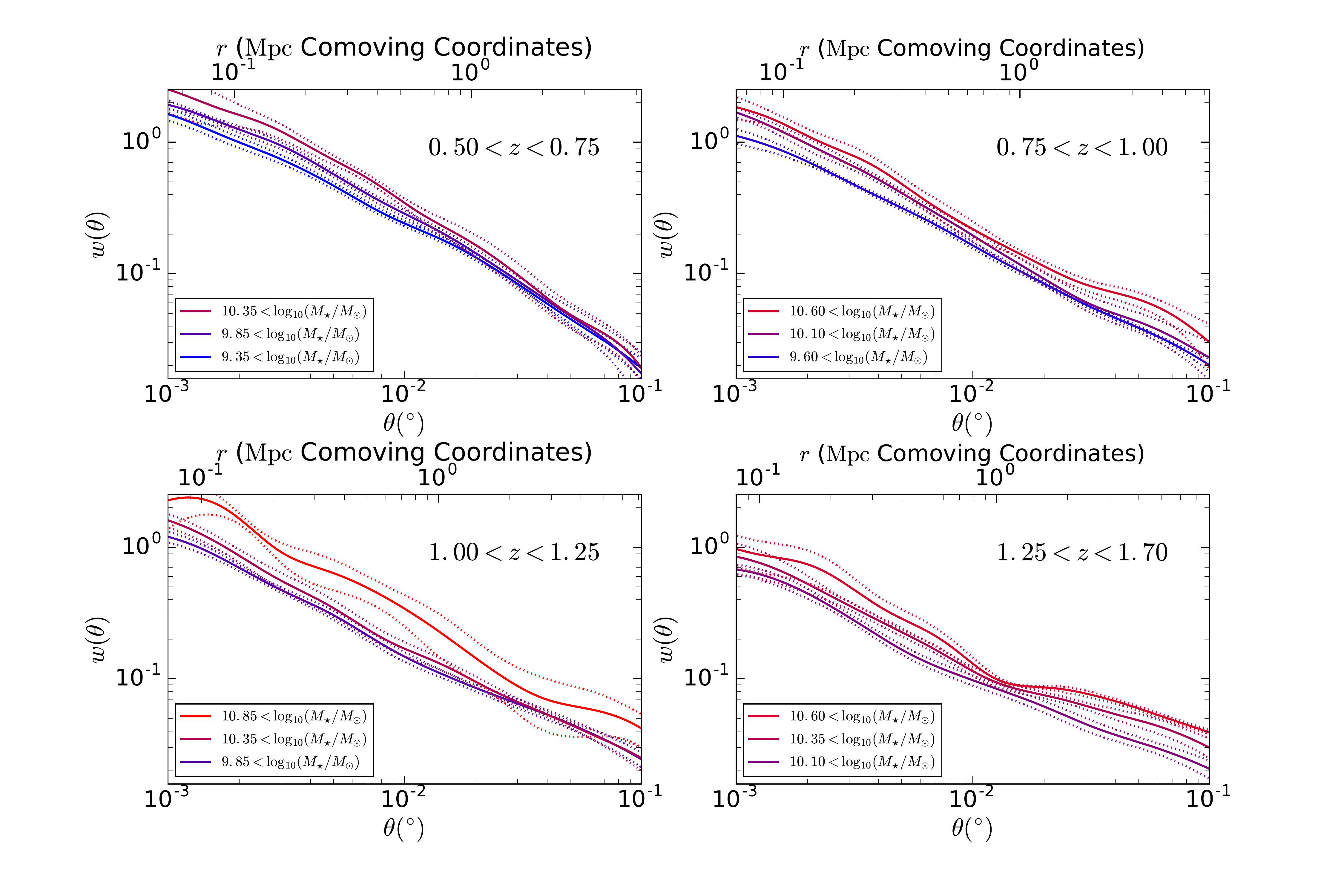}
\caption{The angular correlation function for different redshift ranges and masses denoted in each panel. The lower x-axis denotes angular scale, the upper x-axis the corresponding projected comoving distance, and the y-axis the correlation function. The fainter, dashed, upper and lower bands represent the error bars on the measurements, discussed in section \ref{sec:non_param}. For the clarity of the plot, we only show alternate stellar mass samples in the first three redshift bin subplots.}
\label{fig:ACF_figure_proper}
\end{figure*}

We divided the data into four redshift bins ($0.50<z<0.75$, $0.75<z<1.00$, $1.00<z<1.25$ and $1.25<z<1.70$) and seven mass bins ($10^{9.35}M_{\odot}<M_{\star}$, $10^{9.6}M_{\odot}<M_{\star}$, $10^{9.85}M_{\odot}<M_{\star}$, $10^{10.1}M_{\odot}<M_{\star}$, $10^{10.35}M_{\odot}<M_{\star}$, $10^{10.6}M_{\odot}<M_{\star}$ and $10^{10.85}M_{\odot}<M_{\star}$), see  figure \ref{fig:mass_z} . For each bin, we calculated the angular correlation function as described in section \ref{sec:ACF_def}. Figure \ref{fig:ACF_figure_proper} shows our measurements. In each bin we see near power law behaviour, with some bins suggestive of the kink associated with the transition from the 1-halo to the 2-halo term, although the impact of the integral constraint at large angular scales means this transition is unlikely to be self evident until we have access to the larger angular scales present in the full VIDEO survey. The clear trend of clustering increasing with stellar mass threshold at all scales is visible in all redshift bins. After the correlation function was calculated, we fit a HOD model for each subsample, as described in section \ref{sec:hod_description}.

\begin{figure}
\includegraphics[scale=0.5]{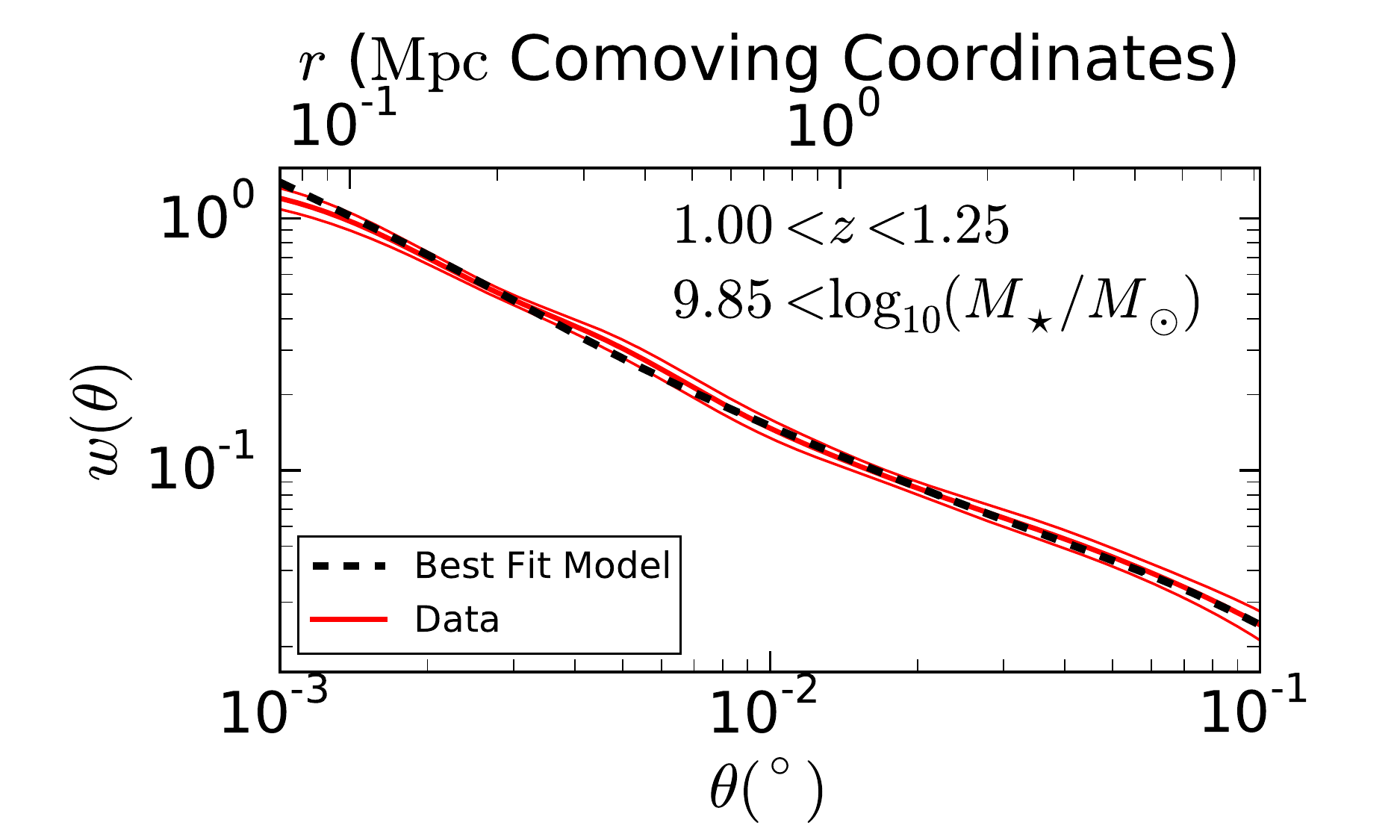}
\caption{Comparison of our best fit model and the clustering data (the band represents the 16th and 84th percentiles of the bootstrapping) for a sample redshift and stellar mass bin. Note that the model is fitted to the number counts as well as the clustering measurements shown here. }
\label{fig:data_model}
\end{figure}

In general, good fits to the data were obtained. Figure \ref{fig:data_model} shows a representative correlation function and best fit model - a close agreement is obtained. Other bins generally achieved similar levels of accord, shown in the appendix in figure \ref{fig:hod_fits_compare}. Our fits had $\chi^2/\textrm{d.o.f.}$ values (showing in table 1) between 0.228 and 1.66, with the exception of the $0.75<z<1$, $10^{10.85}M_{\odot}<M_{\star}$ bin, which had $\chi^2/\textrm{d.o.f.}=4.33$. This suggests the data was well described by the HOD model in all cases apart from one bin. Consistent with these $\chi^2/\textrm{d.o.f.}$ results, the outlier bin in question is seen in figure \ref{fig:hod_fits_compare} to have a correlation function that doesn't fit into the pattern of measurements in the other mass ranges in the same redshift bin - the massive galaxies at this redshift appear to have an unusual spatial distribution (conceivably associated with the sharp peak in the redshift distribution at $z\sim0.8$). Since there was still enough data to make a good measurement of the clustering, and it was possible to make a moderate fit, we still include the results from this bin in subsequent plots, but note the results for this bin might be subject to some unknown systematic effect.

\begin{figure}
\includegraphics[scale=0.5]{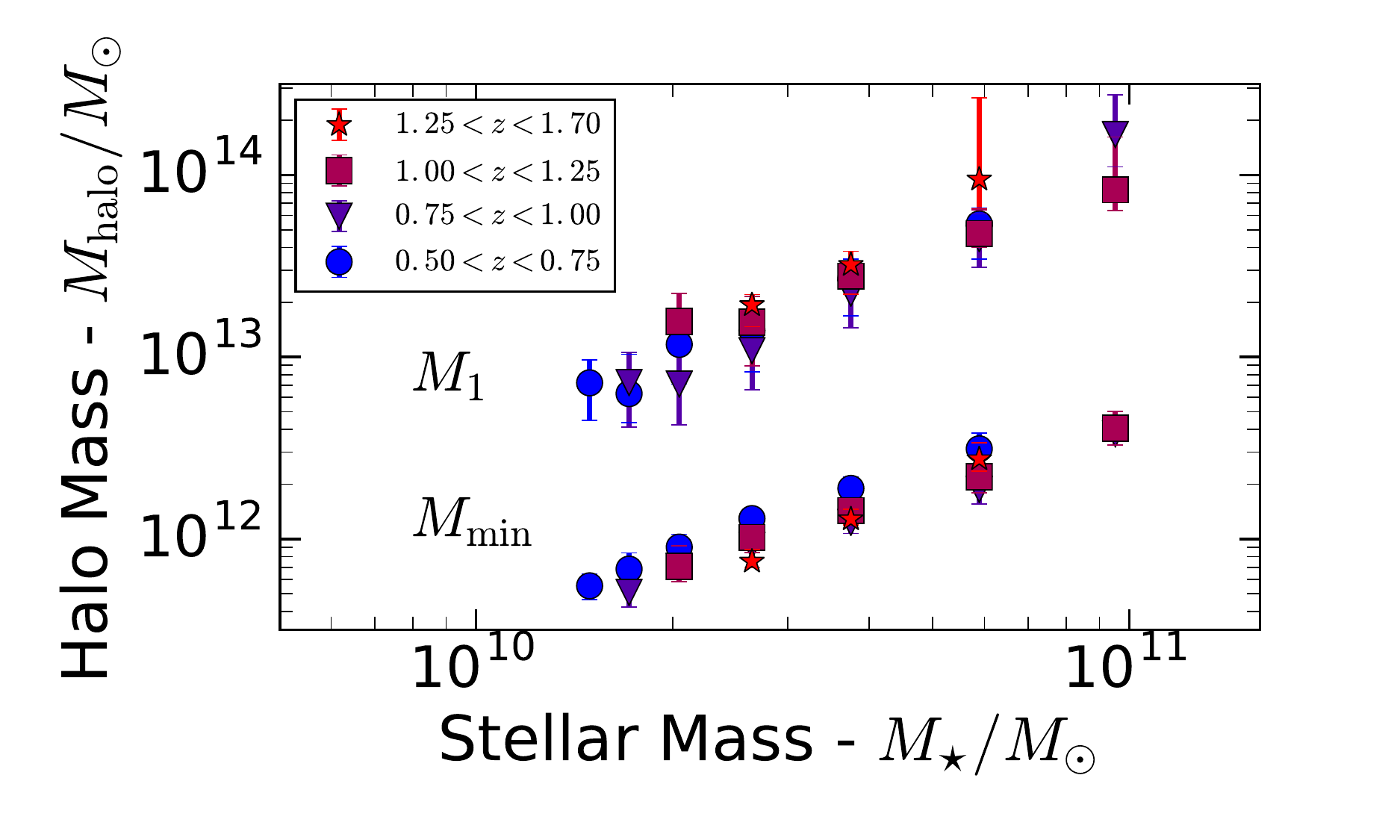}
\caption{Evolution of the HOD parameter $M_{\mathrm{min}}$ and $M_{1}$ as a function of stellar mass in the four redshift bins denoted in the legend.}
\label{fig:hod_1}
\end{figure}

\begin{figure}
\includegraphics[scale=0.5]{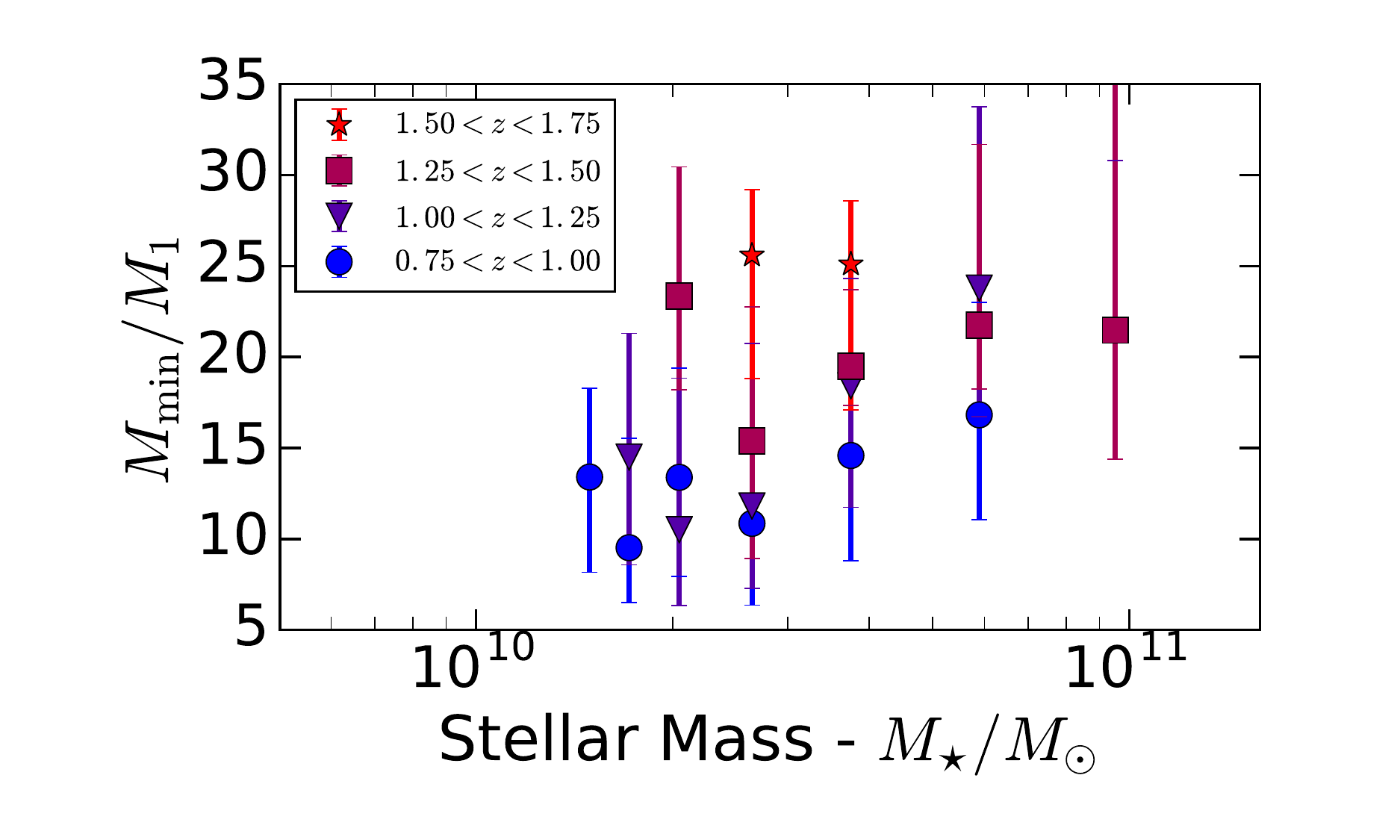}
\caption{Evolution of the ratio of the HOD parameters $M_{1}$ to $M_{\mathrm{min}}$ as a function of stellar mass in the four redshift bins denoted in the legend.}
\label{fig:hod_ratio}
\end{figure}

\begin{figure}
\includegraphics[scale=0.5]{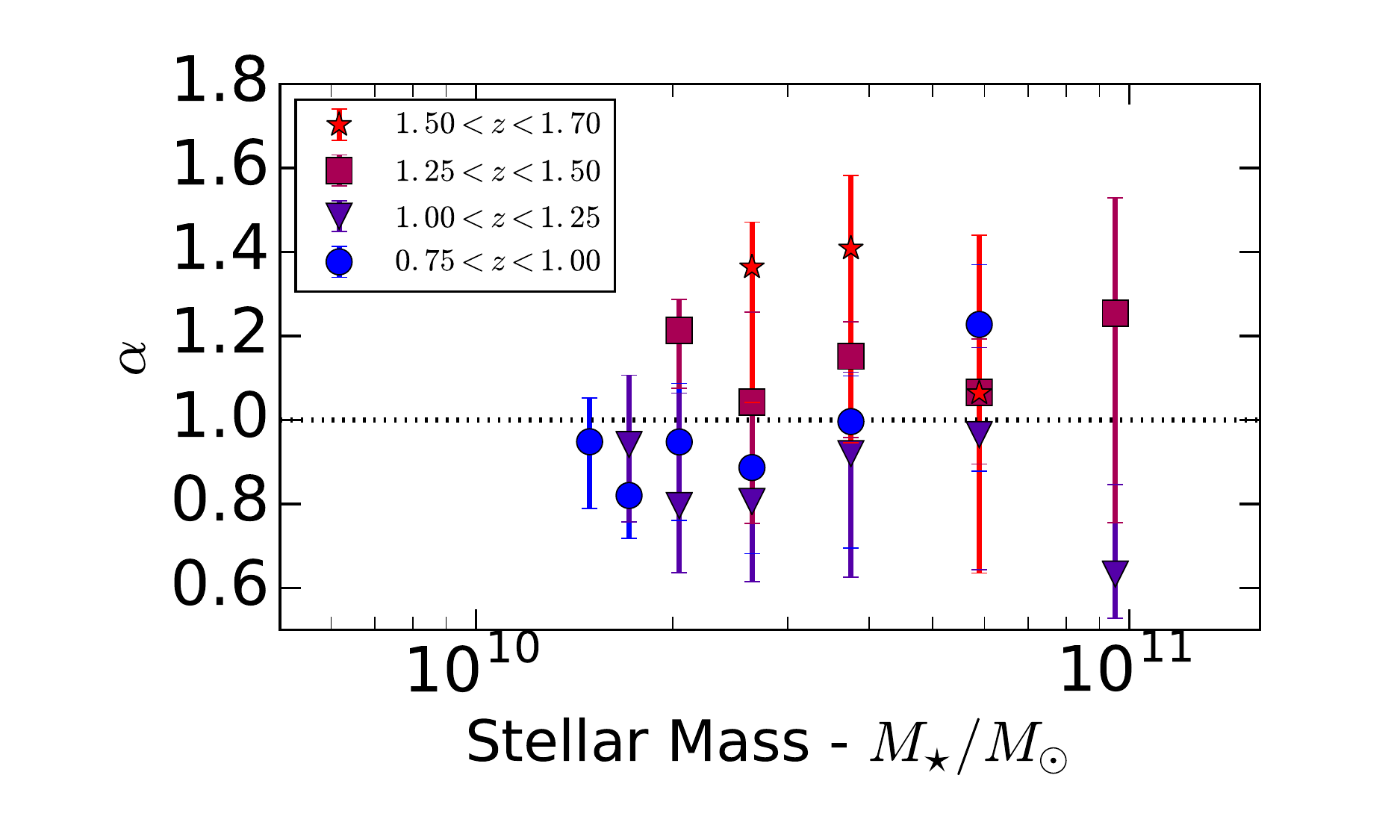}
\caption{Evolution of the HOD parameter $\alpha$ as a function of stellar mass in the four redshift bins denoted in the legend. The dotted line represents $\alpha=1$.}
\label{fig:hod_3}
\end{figure}

\begin{figure}
\includegraphics[scale=0.5]{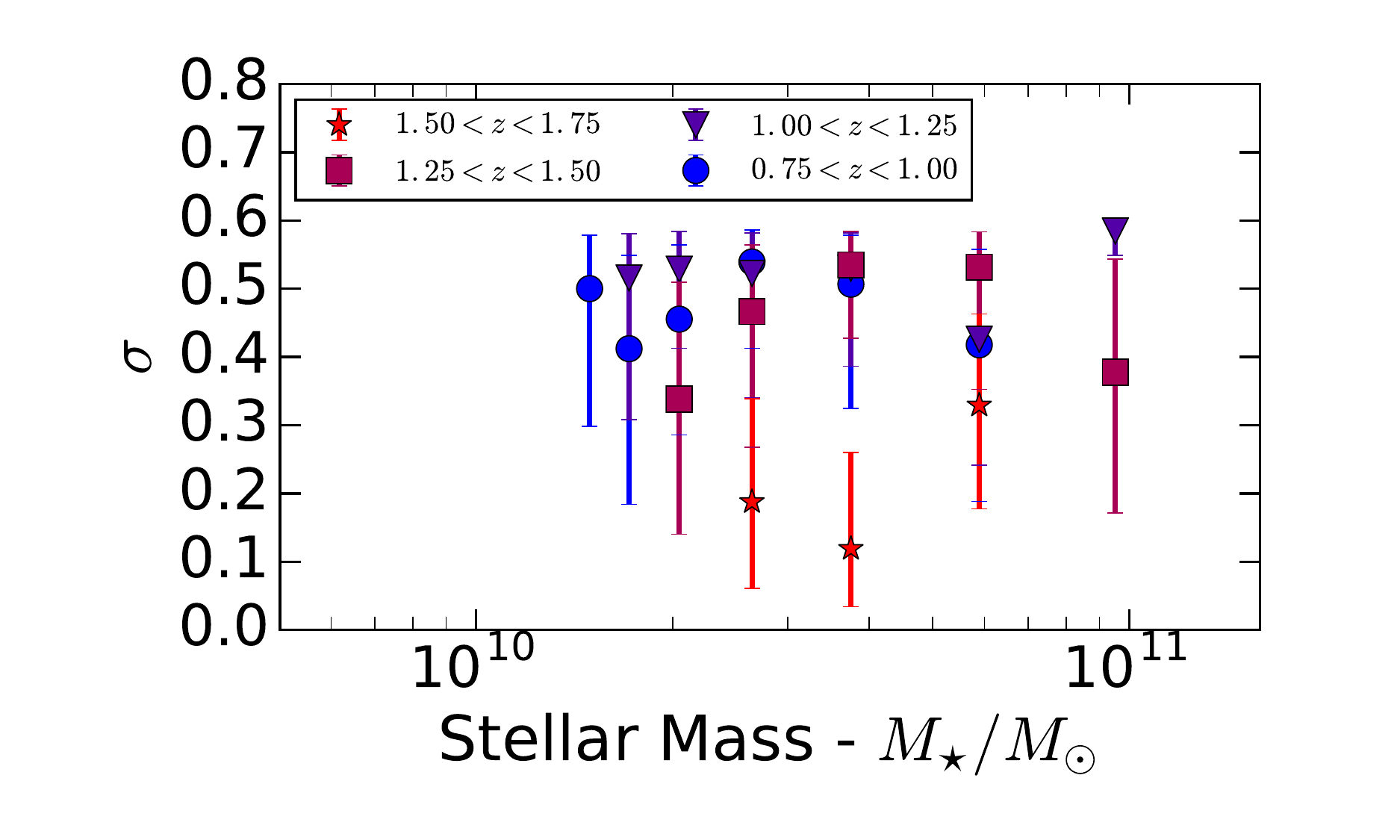}
\caption{Evolution of the HOD parameter $\sigma$ as a function of stellar mass in the four redshift bins denoted in the legend}
\label{fig:hod_2}
\end{figure}

\begin{figure}
\includegraphics[scale=0.5]{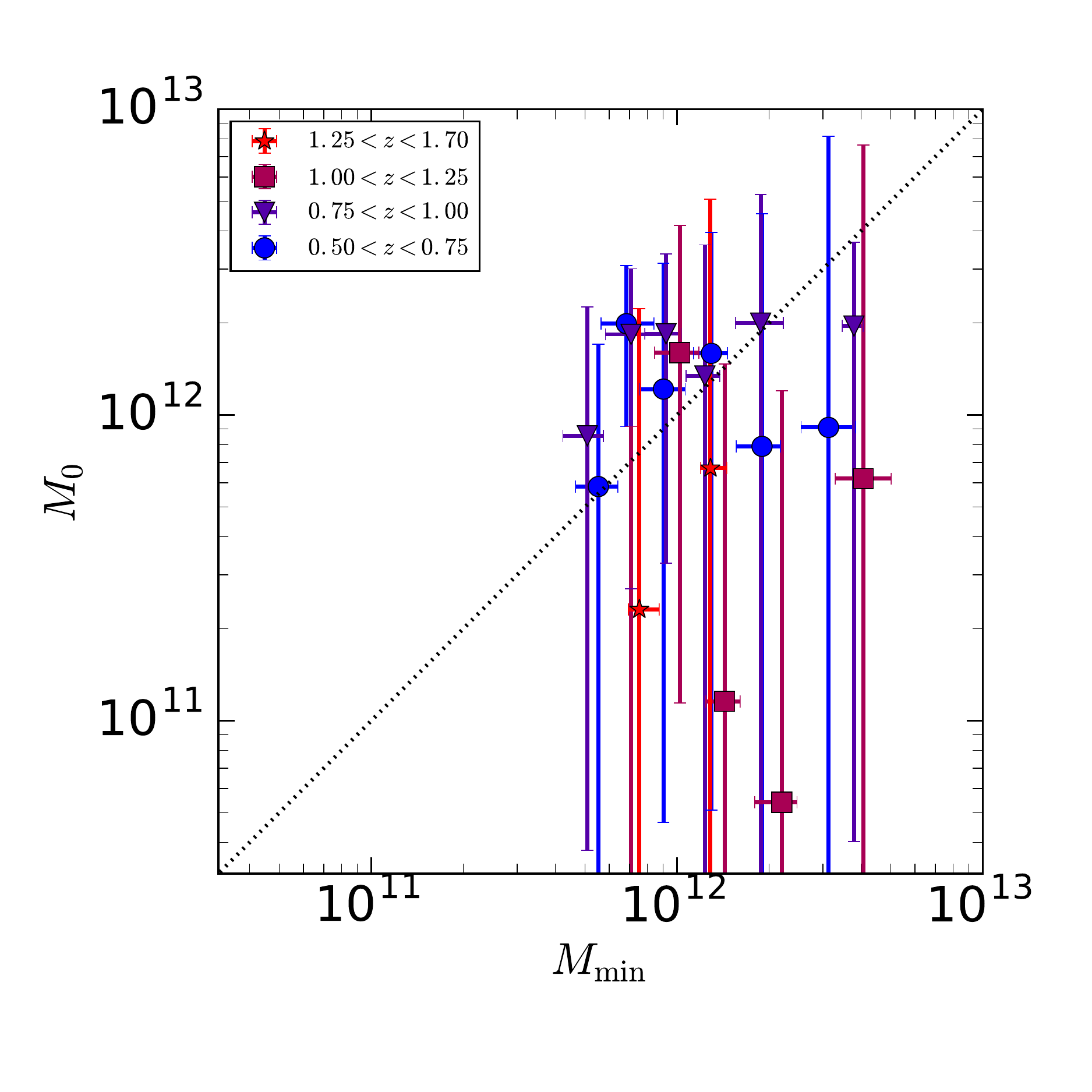}
\caption{Plot comparing HOD parameters $M_{\mathrm{min}}$ and $M_0$ in the four redshift bins denoted in the legend. A one to one line is over-plotted as a guide.}
\label{fig:hod_M0}
\end{figure}

\subsection{$M_{\mathrm{min}}$, $M_{1}$ and the Mass Gap}

Figure \ref{fig:hod_1} shows both $M_{\mathrm{min}}$ and $M_{1}$ growing as approximate power laws with median stellar mass, with little to no evidence of redshift evolution. $M_{\mathrm{min}}$ and $M_{1}$ can be thought of as the halo masses required to host the first and second\footnote{Because of $M_{0}$ this is not strictly true for the second galaxy, however when $M_{0}<<M_{1}$, as for our data, the approximation holds} galaxies in a halo respectively; this shows the well known result that more massive galaxies reside in more massive halos.   $M_1$ remains slightly more than an order of magnitude more massive than $M_{\mathrm{min}}$ over all our stellar masses. We do not detect any upturn in $M_{1}$ at stellar masses $>10^{10.5}\Msol$ seen in \citet{McCracken2015}, although this is perhaps not surprising as we do not reach to as high masses as UltraVISTA (due to the slightly smaller area used in this paper), and based on their results would only expect to see the upturn in our highest mass range.

\begin{figure}
\includegraphics[scale=0.5]{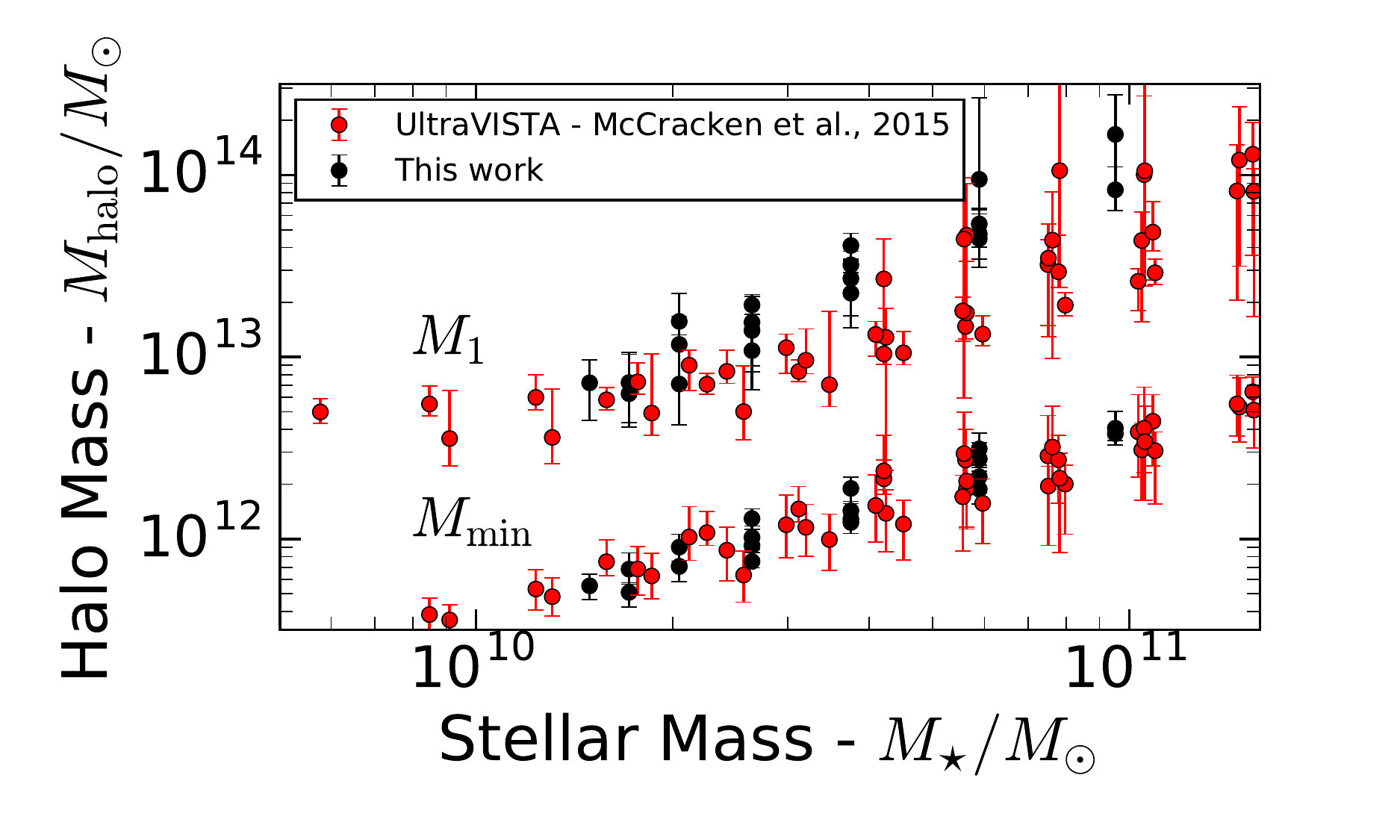}
\caption{Comparison of our HOD results with similar modelling of UltraVISTA data in \citet{McCracken2015}.}
\label{fig:UltraVISTA_comp}
\end{figure}

Figure \ref{fig:hod_ratio} shows $M_{1}/M_{\mathrm{min}}$, the `mass gap' between forming the first galaxy and forming the second. In our stellar mass ranges, the ratio appears to be constant with moderate scatter, potentially increasing at higher stellar masses. Again, we see little redshift evolution, as expected from figure \ref{fig:hod_1}. We do not probe to low enough masses to see if we find an upturn at the lowest masses as in \cite{McCracken2015}. We do find however that this ratio ranges from around 10-20 at these stellar masses (albeit with large error bars), whereas \cite{McCracken2015} finds the ratio to be around 5-10 for the same stellar masses. McCracken suggest that their results help explain why different literature results measure different ratios - that different surveys are biased towards different stellar masses, and thus correctly obtain different results. However this does not explain the discrepancy between our two results, as we are explicitly controlling for stellar mass (although other surveys do report results not dissimilar to those in this paper e.g. \citealp{Zehavi2011}). We suggest the result is likely due to a combination of variance between the fields and the fact that for both VIDEO and UltraVISTA the field sizes only allow access to a small part of the 2-halo term, in the part of angular space where it is most difficult to account for the integral constraint. Figure \ref{fig:UltraVISTA_comp} shows our results plotted alongside the UltraVISTA results, showing close agreement, apart from the high stellar mass end for $M_1$, where we report slightly higher values. VUDS covered both the COSMOS field (that UltraVISTA covers) and largely overlapped the D1 field studied in this paper (in the VUDS VVDS-02h field). \cite{Durkalec2015} report, for samples with otherwise identical selection, slightly more clustering power at larger scales in the COSMOS field, so slight clustering variance between these fields is not without precedence. We anticipate the origin of the discrepancy will become more clear with the full VIDEO survey.

\subsection{$\alpha$ and $\sigma$}

Figure \ref{fig:hod_3} shows $\alpha$ for our four redshift bins, which parametrises how the number of satellites grows with stellar mass. We see very little evolution with either redshift or stellar mass, with typical best fit values of $\sim$1, which can be interpreted as the number of subhalos growing in proportion to the halo halo mass, which is to be expected. Although there appears to be some weak evolution towards higher $\alpha$ values at high redshift, we are cautious to claim a trend for the following reasons. Firstly the trends are of order of the size of our error bars. Secondly, measurements in the same redshift bin are not independent, so trends seen for all the stellar mass values in a given redshift bin relative to another redshift bin are not necessarily significant. Thirdly each redshift bin is a spatially separated part of the Universe, and comparison with UltraVISTA suggests the variation between redshift bins is of a similar order to that expected from cosmic variance. In addition, the five HOD parameters are not independent from each other for a given sample, so samples having outlier values of $\alpha$ can have corresponding outlier values of the other parameters etc.

Figure \ref{fig:hod_2} shows $\sigma$, which parametrises how critical the step jump in halo mass is to form the first galaxy at $M_{\mathrm{min}}$, equivalently the scatter in halo mass at fixed stellar mass for central galaxies. We see no substantial redshift or stellar mass dependence, measuring a constant value of around 0.3-0.5. We note in some samples the posterior of $\sigma$ pushes close to the boundary of our prior. Both \citet{Coupon2012} and \citet{McCracken2015} report similar findings,   \cite{Coupon2012} suggesting sample incompleteness due to photometric errors could lead to missing central galaxies and hence high scatter. Alternatively it could be the case that the $z=0$ motivated 5-parameter model we use here is less appropriate at higher redshifts. \cite{Zheng2005} give an interpretation of $\sigma$ in terms of the scatter in the stellar mass at fixed halo mass: if the functional form for the number of central galaxies is an error function, then at a fixed halo mass the distribution of $\log (M_{\mathrm{gal}})$ is Gaussian. For $M_{\star} \propto M_{\mathrm{\mathrm{halo}}}^{\mu}$ at that halo mass,  the galaxy mass scatter can be expressed by $\mu \times \sigma = \sigma_{M_{\mathrm{gal}}}$, where $\sigma_{M_{\mathrm{gal}}}$ is the scatter in stellar mass at fixed halo mass. Using stellar mass threshold for stellar mass, and $M_{\mathrm{min}}$ for halo mass, (\citealp{Coupon2012}), we measure $\mu \approx 2$ and thus $\sigma_{M_{\mathrm{gal}}} \approx 0.8$. An alternate way of probing the halo mass to stellar mass ratio is the Tully-Fisher relation (\citealt{Tully1977}), which relates rotational velocity to galaxy luminosity for spiral galaxies. However the rotational velocity can give a measure of the dynamical mass, dominated by the halo mass, and modern stellar models can convert the luminosity into a measure of stellar mass (e.g. the redshift and mass according to the VIDEO photometry). \cite{Zheng2005} found a value of $\sigma=0.15$ in their SPH simulation, in good agreement with dispersion measurements in the Tully-Fisher relation at $z=0$. Tiley et al. 2015 (submitted) find a greatly increased scatter at $z\sim1$ relative to $z=0$ in the KMOS Redshift One Survey (KROSS), reporting scatter of 0.32 dex in stellar mass at fixed stellar dynamical mass for their full sample. Although direct comparisons are difficult as the correspondence between dynamical mass and halo mass is non-direct, and the selection methods between wide-field and integral field surveys are very different, the picture of increasing scatter at higher redshifts suggested by the two methods qualitatively agree. Subhalo abundance matching (SHAM) techniques also report dispersion in the order of about 0.2 at low redshift e.g. $z=0.05$ in \cite{Reddick2013}, however, as discussed in that paper, the scatter is partially an underlying assumption of the technique as opposed to a measurement.

\subsection{Insight into Substructure}

\cite{Kravtsov2004} suggest $M_{0}$ can be interpreted in terms of halo substructure. Below this quantity no satellites form, leading to the satellite occupation number to drop off more sharply than a power law. Within the paradigm of satellite galaxies living in subhalos, $M_{0}$ can be viewed as the mass at which the halo is big enough to have enough substructure to have subhalos capable of hosting their own galaxies. Sub-halo abundance matching methods assume that the stellar mass to halo mass ratio is unchanged for sub-halos. If we assume that here, $M_{\mathrm{min}}$ can be viewed as the typical halo mass for that stellar mass, and  $M_{0}$ can be viewed as the minimum halo mass to have a sub halo of mass $M_{\mathrm{min}}$ (and hence a satellite galaxy of that stellar mass). We compare them in figure \ref{fig:hod_M0}, which shows that our constraints on $M_{0}$ are in general poor, but that $M_{0}$ is typically only slightly larger than $M_{\mathrm{min}}$, suggesting that at these times halos were rich in substructure, and could have sub-halos close in size to the whole halo. Conversely, a value of $M_{0}$ significantly larger than $M_{\mathrm{min}}$ would have suggested that halos in the given epoch were poor in substructure, and that all subhalos were dramatically smaller than the mass of the whole halo. This is in contrast to measurements at $z=0$, by which time much of this substructure is destroyed by tidal stripping, and dynamical friction has slowed their orbits until they fall into the centre, e.g. \cite{Zentner2005}, and we see a $M_{0}$ much larger than $M_{\mathrm{min}}$.

\subsection{Derived Parameters} \label{sec:RESULTS_derived}

\begin{figure}
\includegraphics[scale=0.5]{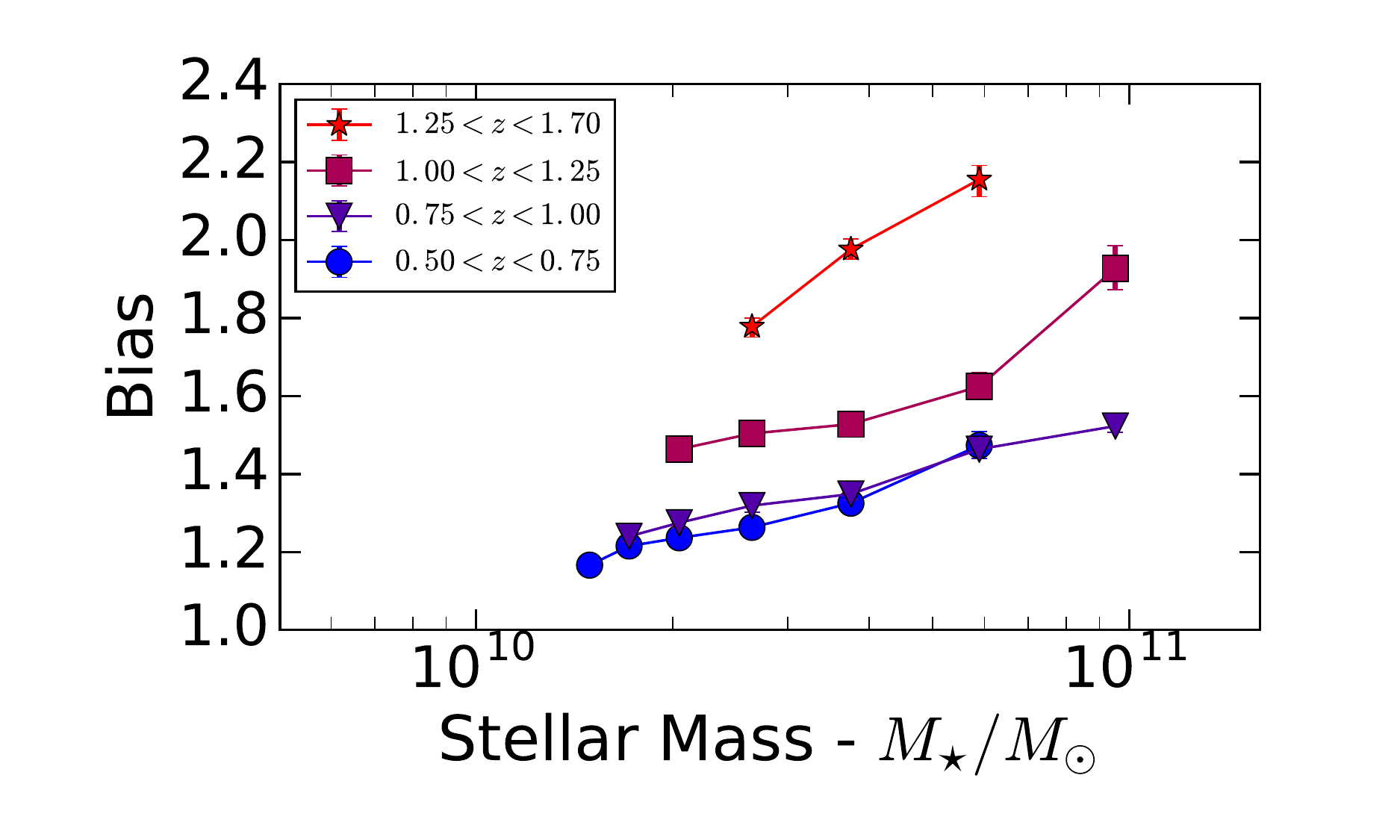}
\caption{Galaxy bias as a function of stellar mass at the four redshift bins denoted in the legend. We see bias increasing with stellar mass, and also increasing with redshift}
\label{fig:bias_evolve}
\end{figure}

We show our measurements of the bias $b$ in figure \ref{fig:bias_evolve} and see the clear decrease in $b$ towards low-$z$ (as galaxies become better tracers of the underlying dark matter and more closely follow its large scale distribution). The established trend of bias increasing with galaxy mass is also evident; corresponding to more massive galaxies preferentially forming in larger dark matter halos, which are themselves more highly bias towards denser regions of dark matter. Note that a decrease in bias with time alone is difficult to interpret. It could be from galaxy populations moving to less massive, less biased halos over time, shifting the median host halo mass to lower values. Or it could be merely from the halo bias evolution. Only with a full HOD analysis can we see that the decreasing bias is predominantly from the latter, that bias-redshift trends are driven by the halos becoming better tracers of the overall matter distribution, as opposed to significant evolution in the galaxy-halo relation.\cite{Durkalec2015} measure the galaxy bias from clustering to be  $\sim$2.6 at $z\sim3$ in the COSMOS field in VUDS, which appears consistent with what one could expect extrapolating our bias measurements to higher redshifts.

\begin{figure}
\includegraphics[scale=0.5]{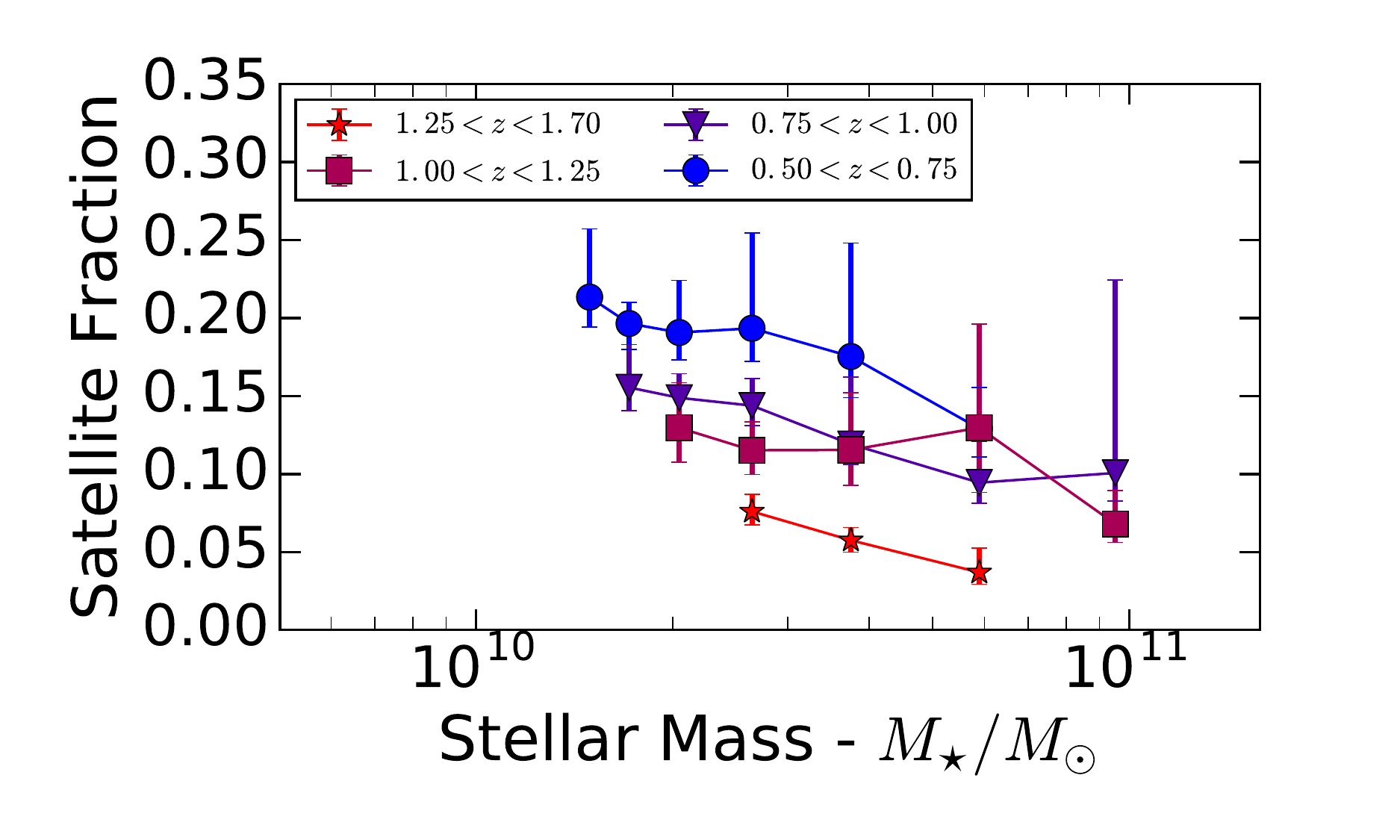}
\caption{The satellite fraction as a function of stellar mass at the four redshift bins denoted in the legend. We see a flat fraction, tailing off at high masses, and more satellites at low $z$ as expected.}
\label{fig:hod_6}
\end{figure}

Our estimates of the satellite fraction (fig. \ref{fig:hod_6}) are consistent with \citet{McCracken2015}, decreasing with redshift (as the higher mass halos within which most satellites reside would be have yet to form). As expected, there are more satellites at low redshifts; at $z \sim1.5$ only around 5\% of galaxies are satellites, which rises to around 20\% by $z \sim0.65$. The satellite fraction also begins to drop off slightly at stellar masses above $ \sim 10^{10.5} M_{\odot}$, as these galaxies are only formed as centrals in higher mass halos, and the even more massive halos within which they would be satellites are very rare.

\begin{figure}
\includegraphics[scale=0.5]{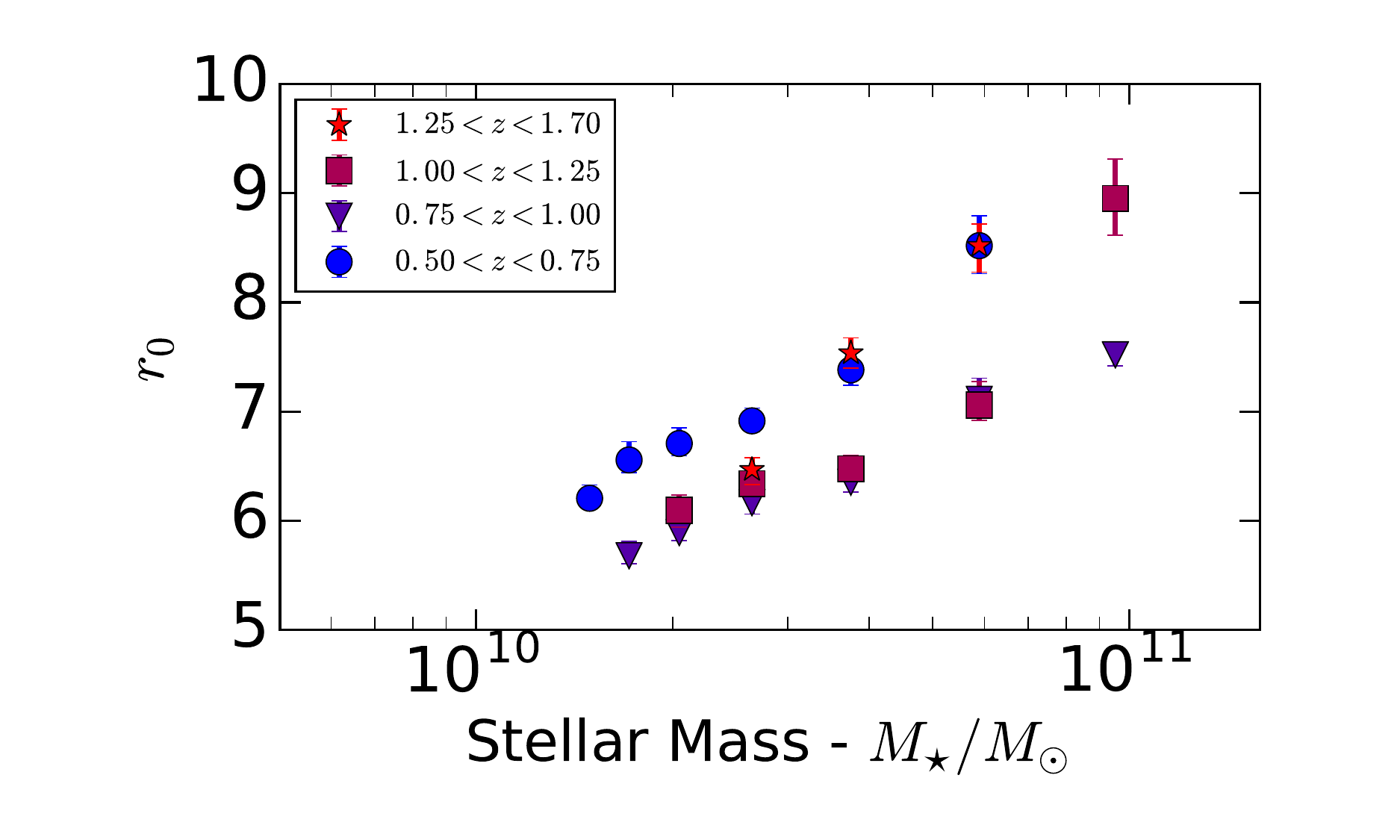}
\caption{Evolution of $r_0$ with stellar mass. Colour represents redshift bin.}
\label{fig:hod_4}
\end{figure}

Our measurements of $r_0$ (figure \ref{fig:hod_4}) are also qualitatively similar to \citet{McCracken2015}, although as with our halo masses we do not yet see evidence for the upturn at higher masses. Our lowest redshift bin is offset relative to the others, but this is to be expected, as it was also offset to higher halo mass values (figure \ref{fig:hod_1}), and even for just fixed halo masses we would expect some $r_0$ evolution as the correlation function of dark matter increases towards lower redshift at fixed radii, independently of the halo occupation.

\subsection{Stellar Mass to Halo Mass Ratio}

\begin{figure}
\includegraphics[scale=0.5]{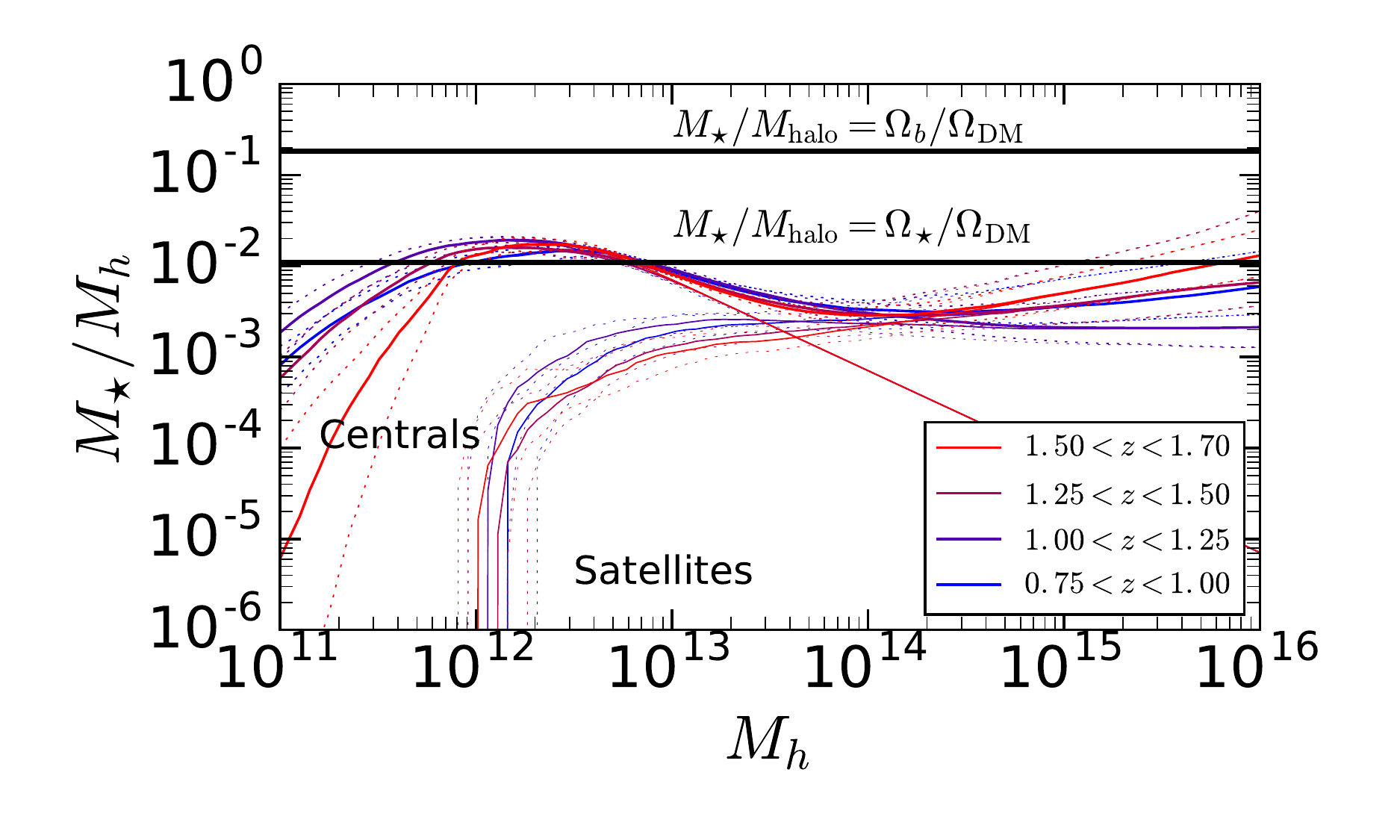}
\caption{The stellar-mass halo ratio for different redshifts, divided into contribution from central galaxy and satellite galaxy. We plot for comparison  $M_{\star} / M_{\mathrm{\mathrm{halo}}}=(\Omega_{b} / \Omega_{\mathrm{DM}}) $, and  $M_{\star} / M_{\mathrm{\mathrm{halo}}}=(\Omega_{\star} / \Omega_{\mathrm{DM}}) $ from \protect\cite{Fukugita2004}}
\label{fig:SMHR}
\end{figure}

\begin{figure}
\includegraphics[scale=0.5]{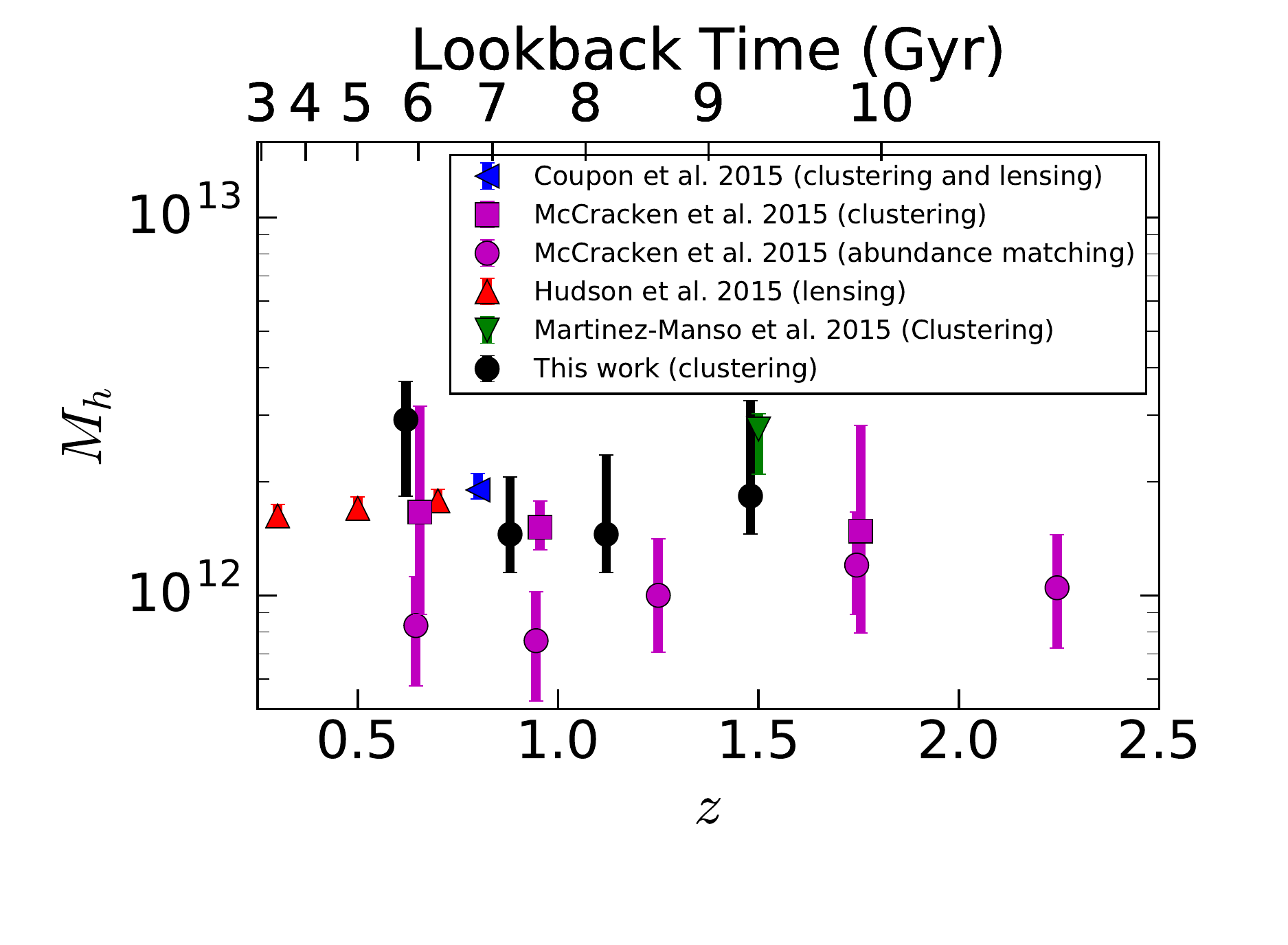}
\caption{The redshift dependence of the peak of the SMHR from the data}
\label{fig:hod_5}
\end{figure}

The stellar mass to halo mass ratio (SMHR) is the total stellar mass in a halo (e.g. the stellar masses of all the galaxies in a halo summed) divided by the halo mass, and can be thought of as a measure of the star-formation and galaxy accretion history of a halo, or it's global star accumulation efficiency for the whole halo. Evidence from the literature suggests it has a peak at halo masses of $10^{11.8}-10^{12.4} M_{\odot}$ (see fig. 11 in \citealp{McCracken2015}, highlighting debate in the literature about possible redshift dependence), and is often modelled with a double power law as in \cite{Yang2003}. We estimate the SMHR by simply integrating our HOD models (broken into central and satellite contributions as in \citealt{Coupon2015}) and bootstrapping the errors (figure \ref{fig:SMHR}). It can also be estimated by analytic inversion if the HOD model is fitted globally (e.g. HOD parameters are expressed as functions of stellar mass, as in \citealt{Coupon2015}) or abundance matching techniques with an N-body simulation (\citealt{Kravtsov2004}, \citealt{Vale2006} and \citealt{Conroy2006}), or from lensing measurements (as in \citealt{Hudson2014}). Simply integrating the HODs suffers systematics in that it can only underestimate the total stellar mass (as you simply do not include lower-mass galaxies that the survey can not detect and it has an artificial upper limit on galaxy mass because the massive galaxies are too rare to make clustering measurements). We also underestimate our error bars by bootstrapping each of our HOD models independently; in practice there is moderate covariance between the models as galaxies appear in multiple correlation functions as we used stellar mass thresholds. However it has the advantage of simplicity and does not make extrapolations to galaxies the survey cannot study - either because of flux or volume limitations. Our estimates of the SMHR rise from very low values at low halo masses (where halos only host a galaxy with a low probability), to reach a peak of $M_{\star}/M_{h} \sim 10^{-1.9}$at a halo mass of $\sim 2  \times10^{12} M_{\odot}$, in a regime where the central galaxy is much more massive than any satellites. It then declines to a local minimum at $\sim 3  \times10^{13} M_{\star}/ M_{\odot}$, where the transition from most stellar mass being in the central galaxy, to most being in satellites, occurs. Subsequently the number of satellites grows as a power law, and the SMHR grows again in the regime of clusters of hundreds of galaxies. This picture is in qualitative agreement (peak and local minimum before a power-law at ultra-massive halos) with \cite{Coupon2015} who conclude that including satellites in the SMHR can boost its value by an order of magnitude, which agrees with our findings. \cite{Durkalec2015a} fit HOD models to the projected correlation function in VUDS and measure the SMHR at $z\sim3$, finding the SMHR reaches the slightly higher value of $M_{\star}/M_{h} \sim 10^{-1.6}$. This could represent slight evolution, although it is very hard to make direct comparisons, given difficulties of making consistent comparisons of stellar mass estimates at different redshifts etc. Our values of halo mass for peak SMHR are consistent with most of the literature values, with weak/no strong redshift dependence (figure \ref{fig:hod_5}). Note that not all techniques of calculating the SMHR are equivalent. Different definitions of halo mass aside ($M_{vir}$ vs. $M_{200}$ etc.), some authors effectively quote the central to halo mass, some (as we do here) quote the sum of the stellar mass of all the galaxies in the halo. \cite{McCracken2015} use median stellar mass to halo mass $M_{\mathrm{min}}$. We, as per \cite{Coupon2012}, use threshold stellar mass to $M_{\mathrm{min}}$, which explains why our SMHR measurements are consistently lower than McCracken. We can be moderately confident in our measurements around the peak, as it is in the range of halo masses probed by our HOD analysis (see fig. \ref{fig:hod_1}), but the estimates far outside this range should be approached with some caution. In particular, the apparent redshift dependence at low stellar masses is an unphysical consequence of not probing to lower stellar masses at high redshifts, and the central galaxy mass to halo mass ratio at high halo masses is unrealistically shallow as we cannot measure clustering for the most massive galaxies.

We also show in figure \ref{fig:SMHR} the line  $M_{\star} / M_{\mathrm{\mathrm{halo}}}=(\Omega_{b} / \Omega_{\mathrm{DM}}) $ (using $\Omega_{b}=0.049$), showing the ratio of baryonic to dark matter for the whole Universe. Our measurements are safely under the line (e.g. we do not have more stellar mass than total baryons!) and indeed show that only a small fraction of baryons are in galaxies, as expected. We also plot the $z=0$ line for the mass estimated to be in stars (6\% of baryons, from \citealp{Fukugita2004}). Although just an average, we see that around the peak of the SMHR, where the main contribution to stellar mass is, we are very close to $\Omega_{\star}/\Omega_{\mathrm{DM}}$, consistent with most, but not all stellar mass seen today having assembled at $z\sim1$.

It is perhaps surprising that we see little evolution in the connection between galaxies stellar mass and their host halos over their redshifts, particularly when other properties of galaxies are known to vary dramatically over the same epoch. For example, global comoving star formation density is well known to drop by around half an order of magnitude from $z\sim2$ to $z\sim0.5$ as galaxies increasingly have less gas to form new stars from, see \citet{Madau2014} for a review. Similarly, the (potentially associated) global super massive black hole accretion density also drops by up to an order of magnitude e.g. \citet{Hirschmann2014}. Morphological properties (e.g. Sersic indices) of galaxies are believed to be relatively stable from the local Universe to $z\sim1$ (\citealp{Cassata2007}), but are typically observed to be dramatically different by $z\sim2$ onwards (\citealp{Lee2013}). All these additional properties are extremely important in understanding galaxy evolution and the galaxy-halo connection. In a follow-up paper we will build on our work here which only incorporates stellar mass measurements, by investigating the interplay between the halo mass and the onset of star formation, incorporating star formation rate measurements and other galaxy properties.

\section{Conclusions} \label{sec:conclusions}

We have used data from the VIDEO survey to investigate the galaxy-halo relation using 10-band photometric redshifts and stellar mass estimates. In particular we have studied the clustering of galaxies with the two point correlation function up to $z\sim1.75$,  using the Parzen-Rosenblatt estimator to calculate the correlation function in a novel way without angular space binning and showing it to be consistent with previous methods.  Then a HOD analysis of the galaxy clustering was performed to give information about how galaxies occupy halos over cosmic time, as well to derive standard properties like the bias of galaxies, and their satellite fraction. On the whole our data was found to be in good agreement with other surveys and clustering analyses, in particular the closely related UltraVISTA survey, another public VISTA survey currently at similar depths and breadths to VIDEO, which in subsequent data releases will get deeper as VIDEO gets wider. 

We see no substantial change in the occupation relations over time; all changes are driven by the change in the halo population. Typical halo mass increases with galaxy mass, and the ratio between the halo being sufficiently massive for one galaxy, to two, is around 15, suggesting that is the typical mass ratio between a halo of a given mass, and a halo massive enough to have enough substructure to have a sub-halo of that given mass. The power law relation for the number of satellites in a halo was $\sim1$, and the scatter in halo mass to galaxy mass broadly consistent with Tully-Fisher measurements at similar redshifts. We found bias increases with stellar mass, as galaxies are found in more massive, more biased halos, and decreases with time, as the halos trace the large scale dark matter distribution more accurately. The  satellite fraction drops at high redshifts as the more massive halo within which satellites are found have not yet collapsed, and at high stellar masses as the super-massive halos within which which high mass galaxies could be satellites are extraordinarily rare. Finally, our estimate of stellar mass to halo mass ratio, although limited by the range of masses VIDEO currently probes, is in reasonable agreement with other studies, with a peak at a halo mass of around $ 2  \times10^{12} M_{\odot}$ that is approximately constant in redshift.

UltraVISTA and VIDEO currently are probing similar parts of parameter space but will start to diverge in future data releases. VIDEO-UltraVISTA complementarity is key - UltraVISTA gives a single instance of structure, we present another here, and future VIDEO results will subsequently give many more. UltraVISTA DR2 will probe several orders of magnitude deeper in the same field in all their near infra-red bands, allowing extension of their analysis to $z>4$, and to lower stellar masses. Future work in subsequent VIDEO releases (when deeper optical data is available) will extend to larger areas over three separate fields (eventually 12deg$^2$ in total), reducing uncertainty on measurements on the parameters reported in this paper, extending the angular scales probed by a factor $\sim$3 (allowing a better constraints on both the 1-halo and 2-halo terms), extending to more massive galaxies (allowing better analysis of the `kink' in halo mass at high stellar masses), and giving an initial measure of cosmic variance by comparing results between the three fields.

\section*{Acknowledgements}

The first author also wishes to acknowledge support provided through an STFC studentship, the Penn State Summer School in Statistics for Astronomers X for introduction to the use of the Parzen and Rozenblatt density estimator, and the Rector and Fellows of Lincoln College for support through the Graduate Research Fund.

AV acknowledges the Leverhulme Trust for support through a Research Fellowship.

The authors thank Steven Murray of the University of Western Australia for the use and advice on his code \textit{halomod} used for the HOD analysis.

Based on data products from observations made with ESO Telescopes at the La Silla or Paranal Observatories under ESO programme ID 179.A- 2006.

Based on observations obtained with MegaPrime/MegaCam, a joint project of CFHT and CEA/IRFU, at the Canada-France-Hawaii Telescope (CFHT) which is operated by the National Research Council (NRC) of Canada, the Institut National des Science de l'Univers of the Centre National de la Recherche Scientifique (CNRS) of France, and the University of Hawaii. This work is based in part on data products produced at Terapix available at the Canadian Astronomy Data Centre as part of the Canada-France-Hawaii Telescope Legacy Survey, a collaborative project of NRC and CNRS.

\bibliographystyle{mn2e_mod}

\bibliography{hatfield_paper}

\begin{thebibliography}{84}
\expandafter\ifx\csname natexlab\endcsname\relax\def\natexlab#1{#1}\fi

\bibitem[{Abbas {et~al}\mbox{.}(2010)Abbas, de~la Torre, {Le F??vre}, Guzzo,
  Marinoni, Meneux, Pollo, Zamorani, Bottini, Garilli, {Le Brun}, Maccagni,
  Scaramella, Scodeggio, Tresse, Vettolani, Zanichelli, Adami, Arnouts,
  Bardelli, Bolzonella, Cappi, Charlot, Ciliegi, Contini, Foucaud, Franzetti,
  Gavignaud, Ilbert, Iovino, Lamareille, McCracken, Marano, Mazure, Merighi,
  Paltani, Pell??, Pozzetti, Radovich, Vergani, Zucca, Bondi, Bongiorno,
  Brinchmann, Cucciati, de~Ravel, Gregorini, Perez-Montero, Mellier, \&
  Merluzzi}]{Abbas2010}
Abbas U. {et~al.}, 2010, MNRAS, 406, 1306

\bibitem[{Arnouts {et~al}\mbox{.}(1999)Arnouts, Cristiani, Moscardini,
  Matarrese, Lucchin, Fontana, \& Giallongo}]{Arnouts1999}
Arnouts S., Cristiani S., Moscardini L., Matarrese S., Lucchin F., Fontana A.,
  Giallongo E., 1999, MNRAS, 310, 540

\bibitem[{Arnouts {et~al}\mbox{.}(2002)Arnouts, Moscardini, Vanzella, Colombi,
  Cristiani, Fontana, Giallongo, Matarrese, \& Saracco}]{Arnouts2002}
Arnouts S. {et~al.}, 2002, MNRAS, 329, 355

\bibitem[{Bahcall \& Soneira(1983)}]{Bahcall1983}
Bahcall N.~A., Soneira R.~M., 1983, ApJ, 270, 20

\bibitem[{Baldry {et~al}\mbox{.}(2010)Baldry, Robotham, Hill, Driver, Liske,
  Norberg, Bamford, Hopkins, Loveday, Peacock, Cameron, Croom, Cross, Doyle,
  Dye, Frenk, Jones, van Kampen, Kelvin, Nichol, Parkinson, Popescu, Prescott,
  Sharp, Sutherland, Thomas, \& Tuffs}]{Baldry2010}
Baldry I.~K. {et~al.}, 2010, MNRAS

\bibitem[{Banerji {et~al}\mbox{.}(2015)Banerji, Jouvel, Lin, McMahon, Lahav,
  Castander, Abdalla, Bertin, Bosman, Carnero, Kind, da~Costa, Gerdes,
  Gschwend, Lima, Maia, Merson, Miller, Ogando, Pellegrini, Reed, Saglia,
  Sanchez, Allam, Annis, Bernstein, Bernstein, Bernstein, Capozzi, Childress,
  Cunha, Davis, DePoy, Desai, Diehl, Doel, Findlay, Finley, Flaugher, Frieman,
  Gaztanaga, Glazebrook, Gonzalez-Fernandez, Gonzalez-Solares, Honscheid,
  Irwin, Jarvis, Kim, Koposov, Kuehn, Kupcu-Yoldas, Lagattuta, Lewis, Lidman,
  Makler, Marriner, Marshall, Miquel, Mohr, Neilsen, Peoples, Sako, Sanchez,
  Scarpine, Schindler, Schubnell, Sevilla, Sharp, Soares-Santos, Swanson,
  Tarle, Thaler, Tucker, Uddin, Wechsler, Wester, Yuan, \& Zuntz}]{Banerji2015}
Banerji M. {et~al.}, 2015, MNRAS, 446, 2523

\bibitem[{Bergvall {et~al}\mbox{.}(2015)Bergvall, Marquart, Way, Blomqvist,
  Holst, {\"{O}}stlin, \& Zackrisson}]{Bergvall2015}
Bergvall N., Marquart T., Way M.~J., Blomqvist A., Holst E., {\"{O}}stlin G.,
  Zackrisson E., 2015, A\&A, 587, A72

\bibitem[{Bertin \& Arnouts(1996)}]{Bertin1996}
Bertin E., Arnouts S., 1996, A\&A Supplement Series, 117, 393

\bibitem[{Beutler {et~al}\mbox{.}(2011)Beutler, Blake, Colless, Jones,
  Staveley-Smith, Campbell, Parker, Saunders, \& Watson}]{Beutler2011}
Beutler F. {et~al.}, 2011, MNRAS, 416, 3017

\bibitem[{Bielby {et~al}\mbox{.}(2014)Bielby, Gonzalez-Perez, McCracken,
  Ilbert, Daddi, {Le F\`{e}vre}, Hudelot, Kneib, Mellier, \&
  Willott}]{Bielby2014}
Bielby R.~M. {et~al.}, 2014, A\&A, 568, A24

\bibitem[{Bowman(1984)}]{Bowman1984}
Bowman A.~W., 1984, Biometrika, 71, 353

\bibitem[{Cassata {et~al}\mbox{.}(2007)Cassata, Guzzo, Franceschini, Scoville,
  Capak, Ellis, Koekemoer, McCracken, Mobasher, Renzini, Ricciardelli,
  Scodeggio, Taniguchi, \& Thompson}]{Cassata2007}
Cassata P. {et~al.}, 2007, ApJ Supplement Series, 172, 270

\bibitem[{Clowes {et~al}\mbox{.}(2013)Clowes, Harris, Raghunathan, Campusano,
  Sochting, \& Graham}]{Clowes2013}
Clowes R.~G., Harris K.~A., Raghunathan S., Campusano L.~E., Sochting I.~K.,
  Graham M.~J., 2013, MNRAS, 429, 2910

\bibitem[{Conroy {et~al}\mbox{.}(2006)Conroy, Wechsler, \&
  Kravtsov}]{Conroy2006}
Conroy C., Wechsler R.~H., Kravtsov A.~V., 2006, ApJ, 647, 201

\bibitem[{Cooray \& Sheth(2002)}]{COORAY2002a}
Cooray A., Sheth R., 2002, Physics Reports, 372, 1

\bibitem[{Coupon {et~al}\mbox{.}(2015)Coupon, Arnouts, van Waerbeke, Moutard,
  Ilbert, van Uitert, Erben, Garilli, Guzzo, Heymans, Hildebrandt, Hoekstra,
  Kilbinger, Kitching, Mellier, Miller, Scodeggio, Bonnett, Branchini,
  Davidzon, {De Lucia}, Fritz, Fu, Hudelot, Hudson, Kuijken, Leauthaud, {Le
  Fevre}, McCracken, Moscardini, Rowe, Schrabback, Semboloni, \&
  Velander}]{Coupon2015}
Coupon J. {et~al.}, 2015, MNRAS, 449, 1352

\bibitem[{Coupon {et~al}\mbox{.}(2012)Coupon, Kilbinger, McCracken, Ilbert,
  Arnouts, Mellier, Abbas, de~la Torre, Goranova, Hudelot, Kneib, \& {Le
  F\`{e}vre}}]{Coupon2012}
Coupon J. {et~al.}, 2012, A\&A, 542, A5

\bibitem[{Davis \& Peebles(1983)}]{Davis1983}
Davis M., Peebles P. J.~E., 1983, ApJ, 267, 465

\bibitem[{Dressler(1980)}]{Dressler1980}
Dressler A., 1980, ApJ, 236, 351

\bibitem[{Driver {et~al}\mbox{.}(2011)Driver, Hill, Kelvin, Robotham, Liske,
  Norberg, Baldry, Bamford, Hopkins, Loveday, Peacock, Andrae, Bland-Hawthorn,
  Brough, Brown, Cameron, Ching, Colless, Conselice, Croom, Cross, {De
  Propris}, Dye, Drinkwater, Ellis, Graham, Grootes, Gunawardhana, Jones, van
  Kampen, Maraston, Nichol, Parkinson, Phillipps, Pimbblet, Popescu, Prescott,
  Roseboom, Sadler, Sansom, Sharp, Smith, Taylor, Thomas, Tuffs, Wijesinghe,
  Dunne, Frenk, Jarvis, Madore, Meyer, Seibert, Staveley-Smith, Sutherland, \&
  Warren}]{Driver2011}
Driver S.~P. {et~al.}, 2011, MNRAS, 413, 971

\bibitem[{Dubois {et~al}\mbox{.}(2014)Dubois, Pichon, Welker, {Le Borgne},
  Devriendt, Laigle, Codis, Pogosyan, Arnouts, Benabed, Bertin, Blaizot,
  Bouchet, Cardoso, Colombi, de~Lapparent, Desjacques, Gavazzi, Kassin, Kimm,
  McCracken, Milliard, Peirani, Prunet, Rouberol, Silk, Slyz, Sousbie,
  Teyssier, Tresse, Treyer, Vibert, \& Volonteri}]{Dubois2014}
Dubois Y. {et~al.}, 2014, MNRAS, 444, 1453

\bibitem[{Durkalec {et~al}\mbox{.}(2015{\natexlab{a}})Durkalec, {Le
  F{\`{e}}vre}, de~la Torre, Pollo, Cassata, Garilli, {Le Brun}, Lemaux,
  Maccagni, Pentericci, Tasca, Thomas, Vanzella, Zamorani, Zucca,
  Amor{\'{\i}}n, Bardelli, Cassar{\`{a}}, Castellano, Cimatti, Cucciati,
  Fontana, Giavalisco, Grazian, Hathi, Ilbert, Paltani, Ribeiro, Schaerer,
  Scodeggio, Sommariva, Talia, Tresse, Vergani, Capak, Charlot, Contini, Cuby,
  Dunlop, Fotopoulou, Koekemoer, L{\'{o}}pez-Sanjuan, Mellier, Pforr, Salvato,
  Scoville, Taniguchi, \& Wang}]{Durkalec2015a}
Durkalec A. {et~al.}, 2015{\natexlab{a}}, A\&A, 576, L7

\bibitem[{Durkalec {et~al}\mbox{.}(2015{\natexlab{b}})Durkalec, {Le
  F{\`{e}}vre}, Pollo, de~la Torre, Cassata, Garilli, {Le Brun}, Lemaux,
  Maccagni, Pentericci, Tasca, Thomas, Vanzella, Zamorani, Zucca,
  Amor{\'{\i}}n, Bardelli, Cassar{\`{a}}, Castellano, Cimatti, Cucciati,
  Fontana, Giavalisco, Grazian, Hathi, Ilbert, Paltani, Ribeiro, Schaerer,
  Scodeggio, Sommariva, Talia, Tresse, Vergani, Capak, Charlot, Contini, Cuby,
  Dunlop, Fotopoulou, Koekemoer, L{\'{o}}pez-Sanjuan, Mellier, Pforr, Salvato,
  Scoville, Taniguchi, \& Wang}]{Durkalec2015}
Durkalec A. {et~al.}, 2015{\natexlab{b}}, A\&A, 583, A128

\bibitem[{Fabian(2012)}]{Fabian2012}
Fabian A., 2012, Annual Review of Astronomy and Astrophysics, 50, 455

\bibitem[{Fisher {et~al}\mbox{.}(1994)Fisher, Davis, Strauss, Yahil, \&
  Huchra}]{Fisher1994}
Fisher K.~B., Davis M., Strauss M.~A., Yahil A., Huchra J., 1994, MNRAS, 266

\bibitem[{Foreman-Mackey {et~al}\mbox{.}(2013)Foreman-Mackey, Hogg, Lang, \&
  Goodman}]{Foreman-Mackey2012}
Foreman-Mackey D., Hogg D.~W., Lang D., Goodman J., 2013, Publications of the
  Astronomical Society of the Pacific, 125, 306

\bibitem[{Fukugita \& Peebles(2004)}]{Fukugita2004}
Fukugita M., Peebles P. J.~E., 2004, ApJ, 616, 643

\bibitem[{Groth \& Peebles(1977)}]{Groth1977}
Groth E.~J., Peebles P. J.~E., 1977, ApJ, 217, 385

\bibitem[{Guo {et~al}\mbox{.}(2015)Guo, Zheng, Behroozi, Zehavi, Chuang,
  Comparat, Favole, Gottloeber, Klypin, Prada, Rodriguez-Torres, Weinberg, \&
  Yepes}]{Guo2015}
Guo H. {et~al.}, 2015, arXiv:1508.07012, 18, 1

\bibitem[{Gwyn(2012)}]{Gwyn2012}
Gwyn S. D.~J., 2012, AJ, 143, 38

\bibitem[{Hartley {et~al}\mbox{.}(2013)Hartley, Almaini, Mortlock, Conselice,
  Grutzbauch, Simpson, Bradshaw, Chuter, Foucaud, Cirasuolo, Dunlop, McLure, \&
  Pearce}]{Hartley2013}
Hartley W.~G. {et~al.}, 2013, MNRAS, 431, 3045

\bibitem[{Hearin {et~al}\mbox{.}(2015)Hearin, Behroozi, \& van~den
  Bosch}]{Hearin2015}
Hearin A.~P., Behroozi P.~S., van~den Bosch F.~C., 2015, arXiv:1504.05578, 0,
  13

\bibitem[{Hirschmann {et~al}\mbox{.}(2014)Hirschmann, {De Lucia}, Wilman,
  Weinmann, Iovino, Cucciati, Zibetti, \& Villalobos}]{Hirschmann2014}
Hirschmann M., {De Lucia} G., Wilman D., Weinmann S., Iovino A., Cucciati O.,
  Zibetti S., Villalobos {\'{A}}., 2014, MNRAS, 23, 24

\bibitem[{Hudson {et~al}\mbox{.}(2014)Hudson, Harris, \& Harris}]{Hudson2014}
Hudson M.~J., Harris G.~L., Harris W.~E., 2014, ApJ, 787, L5

\bibitem[{Ilbert {et~al}\mbox{.}(2006)Ilbert, Arnouts, McCracken, Bolzonella,
  Bertin, {Le F\`{e}vre}, Mellier, Zamorani, Pell\`{o}, Iovino, Tresse, {Le
  Brun}, Bottini, Garilli, Maccagni, Picat, Scaramella, Scodeggio, Vettolani,
  Zanichelli, Adami, Bardelli, Cappi, Charlot, Ciliegi, Contini, Cucciati,
  Foucaud, Franzetti, Gavignaud, Guzzo, Marano, Marinoni, Mazure, Meneux,
  Merighi, Paltani, Pollo, Pozzetti, Radovich, Zucca, Bondi, Bongiorno,
  Busarello, {De La Torre}, Gregorini, Lamareille, Mathez, Merluzzi, Ripepi,
  Rizzo, \& Vergani}]{Ilbert2006}
Ilbert O. {et~al.}, 2006, A\&A, 457, 841

\bibitem[{Jarvis {et~al}\mbox{.}(2013)Jarvis, Bonfield, Bruce, Geach, McAlpine,
  McLure, Gonzalez-Solares, Irwin, Lewis, Yoldas, Andreon, Cross, Emerson,
  Dalton, Dunlop, Hodgkin, Le, Karouzos, Meisenheimer, Oliver, Rawlings,
  Simpson, Smail, Smith, Sullivan, Sutherland, White, \& Zwart}]{Jarvis2013}
Jarvis M.~J. {et~al.}, 2013, MNRAS, 428, 1281

\bibitem[{Johnston {et~al}\mbox{.}(2015)Johnston, Vaccari, Jarvis, Smith,
  Giovannoli, H\"{a}u\ss~ler, \& Prescott}]{Johnston2015}
Johnston R., Vaccari M., Jarvis M., Smith M., Giovannoli E., H\"{a}u\ss~ler B.,
  Prescott M., 2015, MNRAS, 453, 2541

\bibitem[{Kauffmann {et~al}\mbox{.}(2013)Kauffmann, Li, Zhang, \&
  Weinmann}]{Kauffmann2013}
Kauffmann G., Li C., Zhang W., Weinmann S., 2013, MNRAS, 430, 1447

\bibitem[{Kerscher {et~al}\mbox{.}(2000)Kerscher, Szapudi, \&
  Szalay}]{Kerscher2000}
Kerscher M., Szapudi I., Szalay A.~S., 2000, ApJ, 535, L13

\bibitem[{Kravtsov {et~al}\mbox{.}(2004)Kravtsov, Berlind, Wechsler, Klypin,
  Gottlober, Allgood, \& Primack}]{Kravtsov2004}
Kravtsov A.~V., Berlind A.~A., Wechsler R.~H., Klypin A.~A., Gottlober S.,
  Allgood B., Primack J.~R., 2004, ApJ, 609, 35

\bibitem[{Landy \& Szalay(1993)}]{Landy1993}
Landy S.~D., Szalay A.~S., 1993, ApJ, 412, 64

\bibitem[{{Le F{\`{e}}vre} {et~al}\mbox{.}(2013){Le F{\`{e}}vre}, Cassata,
  Cucciati, Garilli, Ilbert, {Le Brun}, Maccagni, Moreau, Scodeggio, Tresse,
  Zamorani, Adami, Arnouts, Bardelli, Bolzonella, Bondi, Bongiorno, Bottini,
  Cappi, Charlot, Ciliegi, Contini, de~la Torre, Foucaud, Franzetti, Gavignaud,
  Guzzo, Iovino, Lemaux, L{\'{o}}pez-Sanjuan, McCracken, Marano, Marinoni,
  Mazure†, Mellier, Merighi, Merluzzi, Paltani, Pell{\`{o}}, Pollo, Pozzetti,
  Scaramella, Tasca, Vergani, Vettolani, Zanichelli, \& Zucca}]{LeFevre2013}
{Le F{\`{e}}vre} O. {et~al.}, 2013, A\&A, 559, A14

\bibitem[{{Le F{\`{e}}vre} {et~al}\mbox{.}(2015){Le F{\`{e}}vre}, Tasca,
  Cassata, Garilli, {Le Brun}, Maccagni, Pentericci, Thomas, Vanzella,
  Zamorani, Zucca, Amorin, Bardelli, Capak, Cassar{\`{a}}, Castellano, Cimatti,
  Cuby, Cucciati, de~la Torre, Durkalec, Fontana, Giavalisco, Grazian, Hathi,
  Ilbert, Lemaux, Moreau, Paltani, Ribeiro, Salvato, Schaerer, Scodeggio,
  Sommariva, Talia, Taniguchi, Tresse, Vergani, Wang, Charlot, Contini,
  Fotopoulou, L{\'{o}}pez-Sanjuan, Mellier, \& Scoville}]{LeFevre2015}
{Le F{\`{e}}vre} O. {et~al.}, 2015, A\&A, 576, A79

\bibitem[{Lee {et~al}\mbox{.}(2013)Lee, Giavalisco, Williams, Guo, Lotz, {Van
  der Wel}, Ferguson, Faber, Koekemoer, Grogin, Kocevski, Conselice, Wuyts,
  Dekel, Kartaltepe, \& Bell}]{Lee2013}
Lee B. {et~al.}, 2013, ApJ, 774, 47

\bibitem[{Limber(1954)}]{Limber:1954zz}
Limber D.~N., 1954, ApJ, 119, 655

\bibitem[{Lindsay {et~al}\mbox{.}(2014)Lindsay, Jarvis, \&
  McAlpine}]{Lindsay2014}
Lindsay S.~N., Jarvis M.~J., McAlpine K., 2014, MNRAS, 440, 2322

\bibitem[{Ling {et~al}\mbox{.}(1986)Ling, Barrow, \& Frenk}]{Ling1986}
Ling E.~N., Barrow J.~D., Frenk C.~S., 1986, MNRAS, 223, 21P

\bibitem[{Madau \& Dickinson(2014)}]{Madau2014}
Madau P., Dickinson M., 2014, Annual Review of A\&A, 52, 415

\bibitem[{McAlpine {et~al}\mbox{.}(2012)McAlpine, Smith, Jarvis, Bonfield, \&
  Fleuren}]{McAlpine2012}
McAlpine K., Smith D. J.~B., Jarvis M.~J., Bonfield D.~G., Fleuren S., 2012,
  MNRAS, 423, 132

\bibitem[{McCracken {et~al}\mbox{.}(2012)McCracken, Milvang-Jensen, Dunlop,
  Franx, Fynbo, {Le F\`{e}vre}, Holt, Caputi, Goranova, Buitrago, Emerson,
  Freudling, Hudelot, L\'{o}pez-Sanjuan, Magnard, Mellier, M\o~ller, Nilsson,
  Sutherland, Tasca, \& Zabl}]{McCracken2012}
McCracken H.~J. {et~al.}, 2012, A\&A, 544, A156

\bibitem[{McCracken {et~al}\mbox{.}(2015)McCracken, Wolk, Colombi, Kilbinger,
  Ilbert, Peirani, Coupon, Dunlop, Milvang-Jensen, Caputi, Aussel, Bethermin,
  \& {Le Fevre}}]{McCracken2015}
McCracken H.~J. {et~al.}, 2015, MNRAS, 449, 901

\bibitem[{Meneux {et~al}\mbox{.}(2009)Meneux, Guzzo, de~la Torre, Porciani,
  Zamorani, Abbas, Bolzonella, Garilli, Iovino, Pozzetti, Zucca, Lilly, {Le
  F\`{e}vre}, Kneib, Carollo, Contini, Mainieri, Renzini, Scodeggio, Bardelli,
  Bongiorno, Caputi, Coppa, Cucciati, de~Ravel, Franzetti, Kampczyk, Knobel,
  Kova\v{c}, Lamareille, {Le Borgne}, {Le Brun}, Maier, Pell\`{o}, Peng, {Perez
  Montero}, Ricciardelli, Silverman, Tanaka, Tasca, Tresse, Vergani, Bottini,
  Cappi, Cimatti, Cassata, Fumana, Koekemoer, Leauthaud, Maccagni, Marinoni,
  McCracken, Memeo, Oesch, \& Scaramella}]{Meneux2009}
Meneux B. {et~al.}, 2009, A\&A, 505, 463

\bibitem[{Navarro {et~al}\mbox{.}(1997)Navarro, Frenk, \& White}]{Navarro1997}
Navarro J., Frenk C., White S., 1997, ApJ, 490, 493

\bibitem[{Oke \& Gunn(1983)}]{Oke1983}
Oke J.~B., Gunn J.~E., 1983, ApJ, 266, 713

\bibitem[{Park {et~al}\mbox{.}(2001)Park, {Gott III}, \& Choi}]{Park2001}
Park C., {Gott III} J.~R., Choi Y.~J., 2001, ApJ, 553, 33

\bibitem[{Parzen(1962)}]{Parzen1962}
Parzen E., 1962, The Annals of Mathematical Statistics, 33, 1065

\bibitem[{Peacock et al. (2001)}]{Peacock2001}
Peacock J.~A. et al., 2001, Nature, 410, 169

\bibitem[{Peebles(1980)}]{PhillipJamesEdwinP1980}
Peebles P. J.~E., 1980, {The Large Scale Structure of the Universe}

\bibitem[{Peng {et~al}\mbox{.}(2010)Peng, Lilly, Kova\v{c}, Bolzonella,
  Pozzetti, Renzini, Zamorani, Ilbert, Knobel, Iovino, Maier, Cucciati, Tasca,
  Carollo, Silverman, Kampczyk, de~Ravel, Sanders, Scoville, Contini, Mainieri,
  Scodeggio, Kneib, {Le F\`{e}vre}, Bardelli, Bongiorno, Caputi, Coppa, de~la
  Torre, Franzetti, Garilli, Lamareille, {Le Borgne}, {Le Brun}, Mignoli,
  Montero, Pello, Ricciardelli, Tanaka, Tresse, Vergani, Welikala, Zucca,
  Oesch, Abbas, Barnes, Bordoloi, Bottini, Cappi, Cassata, Cimatti, Fumana,
  Hasinger, Koekemoer, Leauthaud, Maccagni, Marinoni, McCracken, Memeo, Meneux,
  Nair, Porciani, Presotto, \& Scaramella}]{Peng2010}
Peng Y.-j. {et~al.}, 2010, ApJ, 721, 193

\bibitem[{Press \& Schechter(1974)}]{Press1974a}
Press W.~H., Schechter P., 1974, ApJ, 187, 425

\bibitem[{Reddick {et~al}\mbox{.}(2013)Reddick, Wechsler, Tinker, \&
  Behroozi}]{Reddick2013}
Reddick R.~M., Wechsler R.~H., Tinker J.~L., Behroozi P.~S., 2013, ApJ, 771, 30

\bibitem[{Roche \& Eales(1999)}]{Roche1999}
Roche N., Eales S.~A., 1999, MNRAS, 307, 703

\bibitem[{Rodriguez {et~al}\mbox{.}(2015)Rodriguez, Merch\'{a}n, \&
  Sgr\'{o}}]{Rodriguez2015}
Rodriguez F., Merch\'{a}n M., Sgr\'{o} M.~A., 2015, A\&A, 580, A86

\bibitem[{Rosenblatt(1956)}]{Rosenblatt1956}
Rosenblatt M., 1956, The Annals of Mathematical Statistics, 27, 832

\bibitem[{Sheth \& Tormen(1999)}]{Sheth1999}
Sheth R.~K., Tormen G., 1999, MNRAS, 308, 119

\bibitem[{Simon {et~al}\mbox{.}(2009)Simon, Hetterscheidt, Wolf, Meisenheimer,
  Hildebrandt, Schneider, Schirmer, \& Erben}]{Simon2008}
Simon P., Hetterscheidt M., Wolf C., Meisenheimer K., Hildebrandt H., Schneider
  P., Schirmer M., Erben T., 2009, MNRAS, 398, 807

\bibitem[{Tinker {et~al}\mbox{.}(2010)Tinker, Robertson, Kravtsov, Klypin,
  Warren, Yepes, Gottlober, \& Gottl{\"{o}}ber}]{Tinker2010}
Tinker J. L.~J., Robertson B. E.~B., Kravtsov a. A.~V., Klypin A., Warren M.
  S.~M., Yepes G., Gottlober S., Gottl{\"{o}}ber S., 2010, ApJ, 724, 11

\bibitem[{Tomczak {et~al}\mbox{.}(2014)Tomczak, Quadri, Tran, Labb{\'{e}},
  Straatman, Papovich, Glazebrook, Allen, Brammer, Kacprzak, Kawinwanichakij,
  Kelson, McCarthy, Mehrtens, Monson, Persson, Spitler, Tilvi, \& van
  Dokkum}]{Tomczak2013}
Tomczak A.~R. {et~al.}, 2014, ApJ, 783, 85

\bibitem[{Trenti \& Stiavelli(2008)}]{Trenti2008}
Trenti M., Stiavelli M., 2008, ApJ, 676, 767

\bibitem[{Tully \& Fisher(1977)}]{Tully1977}
Tully R.~B., Fisher J.~R., 1977, A\&A, 54, 661

\bibitem[{Vale \& Ostriker(2006)}]{Vale2006}
Vale A., Ostriker J.~P., 2006, MNRAS, 371, 1173

\bibitem[{Van~den bosch {et~al}\mbox{.}(2013)Van~den bosch, More, Cacciato, Mo,
  \& Yang}]{Bosch2012}
Van~den bosch F.~C., More S., Cacciato M., Mo H., Yang X., 2013, MNRAS, 430,
  725

\bibitem[{Wang {et~al}\mbox{.}(2006)Wang, Li, Kauffmann, \& {De
  Lucia}}]{Wang2006}
Wang L., Li C., Kauffmann G., {De Lucia} G., 2006, MNRAS, 371, 537

\bibitem[{Warren {et~al}\mbox{.}(2006)Warren, Abazajian, Holz, \&
  Teodoro}]{Warren2006}
Warren M.~S., Abazajian K., Holz D.~E., Teodoro L., 2006, ApJ, 646, 881

\bibitem[{Weinmann {et~al}\mbox{.}(2006)Weinmann, van~den Bosch, Yang, \&
  Mo}]{Weinmann2006}
Weinmann S.~M., van~den Bosch F.~C., Yang X., Mo H.~J., 2006, MNRAS, 366, 2

\bibitem[{White(1979)}]{White1979}
White S., 1979, MNRAS, 186, 145

\bibitem[{White {et~al}\mbox{.}(2015)White, Jarvis, Haussler, \&
  Maddox}]{White2015}
White S.~V., Jarvis M.~J., Haussler B., Maddox N., 2015, MNRAS, 448, 2665

\bibitem[{Willett {et~al}\mbox{.}(2013)Willett, Lintott, Bamford, Masters,
  Simmons, Casteels, Edmondson, Fortson, Kaviraj, Keel, Melvin, Nichol, {Jordan
  Raddick}, Schawinski, Simpson, Skibba, Smith, \& Thomas}]{Willett2013}
Willett K.~W. {et~al.}, 2013, MNRAS, 435, 2835

\bibitem[{Yang {et~al}\mbox{.}(2003)Yang, Mo, \& Bosch}]{Yang2003}
Yang X., Mo H.~J., Bosch F. C. v.~d., 2003, MNRAS, 339, 1057

\bibitem[{Yang {et~al}\mbox{.}(2005)Yang, Mo, Jing, \& van~den
  Bosch}]{Yang2005}
Yang X., Mo H.~J., Jing Y.~P., van~den Bosch F.~C., 2005, MNRAS, 358, 217

\bibitem[{Zehavi {et~al}\mbox{.}(2005)Zehavi, Eisenstein, Nichol, Blanton,
  Hogg, Brinkmann, Loveday, Meiksin, Schneider, \& Tegmark}]{Zehavi2005}
Zehavi I. {et~al.}, 2005, ApJ, 621, 22

\bibitem[{Zehavi {et~al}\mbox{.}(2011)Zehavi, Zheng, Weinberg, Blanton,
  Bahcall, Berlind, Brinkmann, Frieman, Gunn, Lupton, Nichol, Percival,
  Schneider, Skibba, Strauss, Tegmark, \& York}]{Zehavi2011}
Zehavi I. {et~al.}, 2011, ApJ, 736, 59

\bibitem[{Zentner {et~al}\mbox{.}(2005)Zentner, Berlind, Bullock, Kravtsov, \&
  Wechsler}]{Zentner2005}
Zentner A.~R., Berlind A.~A., Bullock J.~S., Kravtsov A.~V., Wechsler R.~H.,
  2005, ApJ, 624, 505

\bibitem[{Zheng {et~al}\mbox{.}(2005)Zheng, Berlind, Weinberg, Benson, Baugh,
  Cole, Dave, Frenk, Katz, \& Lacey}]{Zheng2005}
Zheng Z. {et~al.}, 2005, ApJ, 633, 791

\end{thebibliography}

\appendix

\section[]{HOD Fits} \label{sec:show_HOD_fits}

\begin{figure*}
\includegraphics[scale=0.4]{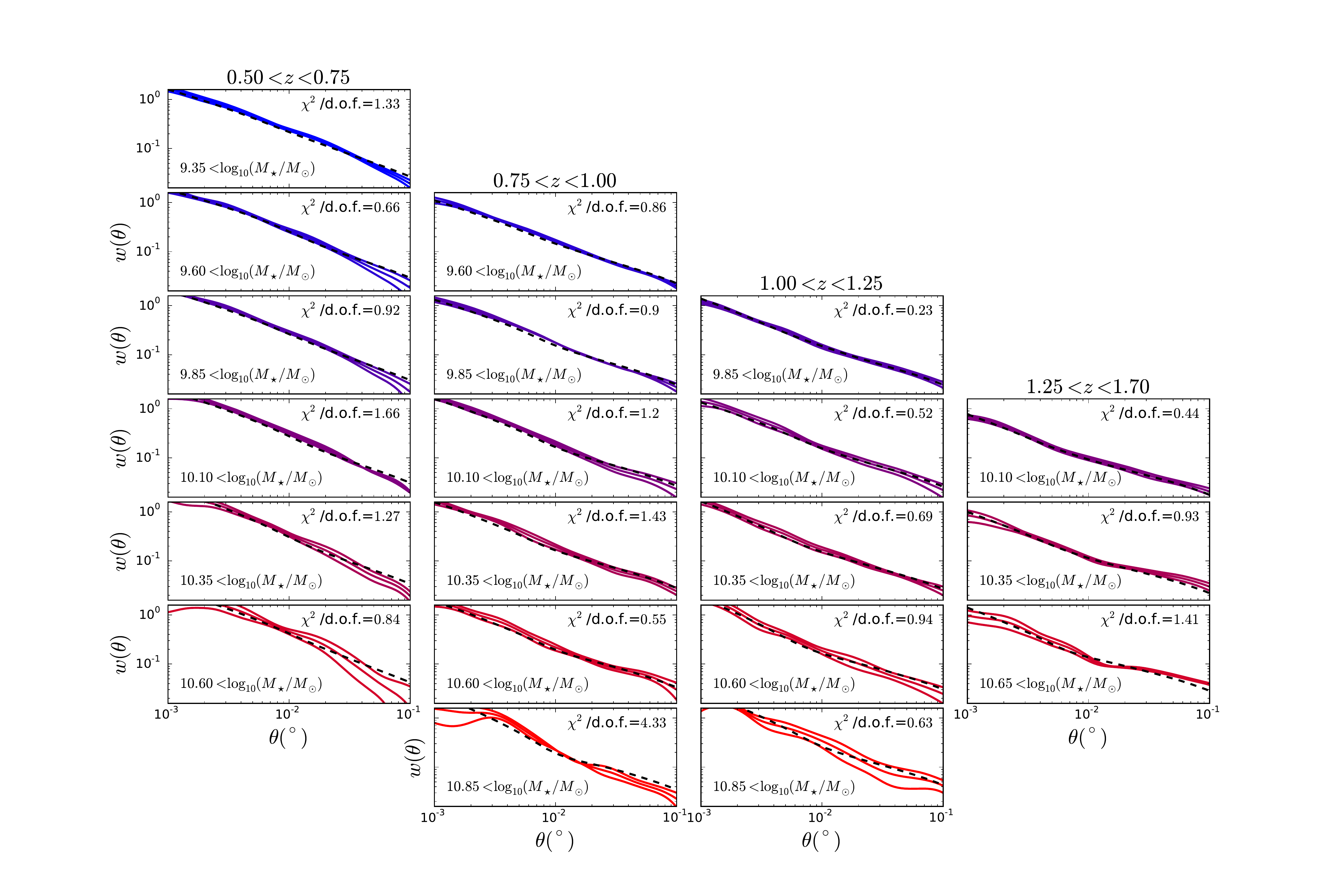}
\caption{The measured correlation functions in our data and the corresponding HOD best fits. Sub-figures in the same column have the same redshift, sub-figures in the same row have the same stellar mass range. The coloured filled lines are the data (blue to red corresponding to increasing stellar mass), and the lower and upper bands are 16th and 84th percentiles from the bootstrapping. The dashed black line is the model correlation function from the best fits. $\chi^2$ values for each fit are shown in the upper right of each sub-figure.}
\label{fig:hod_fits_compare}
\end{figure*}

We show in figure \ref{fig:hod_fits_compare} our measurements and corresponding HOD fits to the data discussed in section \ref{sec:RESULTS}. Good fits were obtained in all cases apart from the $0.75<z<1$, $10^{10.85}M_{\odot}<M_{\star}$ bin.

\bsp

\label{lastpage}

\end{document}